\documentclass[aps,pra,twocolumn,superscriptaddress]{revtex4-2}
\usepackage{graphicx}
\usepackage{dcolumn}
\usepackage{bm}
\usepackage{physics}
\usepackage{siunitx}
\usepackage{amsmath}
\usepackage{hyperref}
\usepackage[export]{adjustbox}
\usepackage{afterpage}
\usepackage{accents}
\usepackage{nccmath}
\usepackage{xcolor}
\usepackage{float} 

\begin{document}

\title{Impact of rotation on a cold atom interferometer and compensation strategy}

\author{Noémie Marquet}
\email[Present address ]{noemie.marquet@lkb.upmc.fr}
\affiliation{DPHY, ONERA, Universit\'{e} Paris-Saclay, 91120 Palaiseau, France}
\author{Yannick Bidel}
\affiliation{DPHY, ONERA, Universit\'{e} Paris-Saclay, 91120 Palaiseau, France}
\author{Malo Cadoret}
\affiliation{LCM, CNAM, 93210 La Plaine Saint-Denis, France}
\author{Alexis Bonnin}
\affiliation{DPHY, ONERA, Universit\'{e} Paris-Saclay, 91120 Palaiseau, France}
\author{Sylvain Schwartz}
\affiliation{DPHY, ONERA, Universit\'{e} Paris-Saclay, 91120 Palaiseau, France}
\author{Phuong-Anh Huynh}
\affiliation{DPHY, ONERA, Universit\'{e} Paris-Saclay, 92320 Châtillon, France}
\author{Alexandre Bresson}
\affiliation{DPHY, ONERA, Universit\'{e} Paris-Saclay, 91120 Palaiseau, France}
\author{Antoine Godard}
\affiliation{ONERA, Universit\'{e} Paris-Saclay, 91120 Palaiseau, France}
\author{Franck Pereira Dos Santos}
\affiliation{Laboratoire Temps Espace, Observatoire de Paris, Universit\'{e} PSL, Sorbonne Universit\'{e}, Universit\'{e} de Lille, LNE, CNRS, 75014 Paris, France}
\author{Olivier Carraz}
\affiliation{ESTEC,ESA, Noordwijk, The Netherlands}
\author{Nassim Zahzam}
\email[]{nassim.zahzam@onera.fr}
\affiliation{DPHY, ONERA, Universit\'{e} Paris-Saclay, 91120 Palaiseau, France}

\date{\today}

\begin{abstract}
Rotations play a detrimental role in achieving ultra-high-performance inertial measurements with an atom interferometer, leading potentially to a total loss of interference contrast and the emergence of dominant phase shift biases. This becomes particularly significant when considering operation in dynamic conditions such as those encountered in Earth orbiting satellites in the perspective of future space gravity missions on-boarding a cold atom accelerometer. We study in this context the impact of rotation on the phase shift and contrast of an atom interferometer and investigate mitigation strategies. An analytical model is derived and compared to experimental demonstrations carried out using an original setup in which the well-controlled proof-mass of a space electrostatic accelerometer is used as the retro-reflection mirror of a cold atom gravimeter. By properly counter-rotating the electrostatic proof-mass, we demonstrate for instance the possibility of recovering the interferometer contrast, otherwise equal to zero, to a level better than 90\%, in both cases of constant angular velocities or in presence of angular accelerations. Our results demonstrate the possibility to perform high performance inertial measurements with a cold atom interferometer in a challenging environments.\\
\end{abstract}

\maketitle
\section*{Introduction}

Atom interferometry addresses the needs for high precision measurements in a wide range of domains, spanning from fundamental physics with for instance the determination of the fine structure constant \citep{morel_determination_2020,parker_measurement_2018}, or with the test of the Weak Equivalence Principle \citep{asenbaum_atom-interferometric_2020}, to more applied domains such as inertial navigation \citep{jekeli_navigation_2005, cheiney_navigation_2018}, or space gravimetry \citep{carraz_spaceborne_2014,chiow_laser-ranging_2015,douch_simulation-based_2018,trimeche_concept_2019,migliaccio_mocass_2019,abrykosov_impact_2019,leveque_gravity_2021,zahzam_hybrid_2022,hosseiniarani_advances_2024}. As this technology gets more mature, inertial sensors such as accelerometers or gyroscopes based on atom interferometry begin to move outside the laboratory. Cold-atom accelerometers have already been used for the last few years for field applications onboard trucks, boats, or planes \citep{geiger_detecting_2011, wu_gravity_2019, bidel_absolute_2018,Wu_Marine_2023, bidel_airborne_2023}. However, such sensors are highly affected by the rotations of the dynamic environment they operate in \citep{bongs_high-order_2006,hogan_light-pulse_2007, lan_influence_2012, barrett_dual_2016,zhao_extension_2021,zahzam_hybrid_2022,beaufils_rotation_2023,darmagnac_de_castanet_atom_2024}, which potentially leads to complete loss of interferometer contrast and significant phase-shift errors. 

To avoid such detrimental impacts of rotation, gyro-stabilized platforms have been efficiently used \citep{bidel_absolute_2018,bidel_airborne_2023}, but this solution comes with a significant increase in instrument size, power consumption and cost, limiting its range of applications. An alternative approach consists in counter-rotating the mirror on which the laser, enabling the atom interferometry measurement, is retro-reflected \citep{lan_influence_2012,hogan_light-pulse_2007}. This mirror can be seen as the reference for the atomic measurement, and maintaining it non-rotating in the inertial frame seems a promising mitigation method. In laboratory environments, this approach was previously implemented using a piezoelectric-driven tip-tilt mirror to prevent Earth's rotation from significantly impacting static cold-atom interferometers \citep{lan_influence_2012,hauth_first_2013,dickerson_multiaxis_2013,duan_suppression_2020}. The same method was also recently demonstrated to show promising results for higher rotation rates up to \SI{250}{\milli \radian \per \s} on a cold-atom interferometer, ultimately dedicated to field applications \citep{darmagnac_de_castanet_atom_2024}.

Overcoming these rotation issues is one of the main challenges to deal with to operate an atom accelerometer under dynamic conditions, particularly in the context of future space gravimetry missions \citep{trimeche_concept_2019,zahzam_hybrid_2022,beaufils_rotation_2023}. Nadin-pointing navigation is then preferred, so the satellite is orbiting around the Earth at a nominal angular velocity of $\approx$ \SI{1.1}{\milli \radian \per \s}. In the absence of rotation mitigation strategies, the contrast and phase shifts of the interferometer are impacted to a level preventing high performances measurements.

In this context, we report here a detailed study analyzing the impact of rotations on a cold-atom accelerometer. This work is conducted through the development of analytical models showing results in terms of contrast and phase for several situations: when the whole instrument is subjected to rotation, when only the interferometer mirror is rotated, and when the rotation compensation strategy is implemented. The results of the models are confronted with experimental studies carried out with an original setup based on a hybrid atomic-electrostatic accelerometer \citep{zahzam_hybrid_2022}, composed of a cold-atom gravimeter associated to a space electrostatic accelerometer (EA) \citep{christophe_new_2015}, adapted to ground operation. In this configuration, the laser used for atom interferometry is directly retro-reflected on the EA's proof-mass acting as the reference mirror for the atom interferometer. The proof-mass position is precisely controlled by use of electrostatic forces allowing its rotation and continuously monitored by capacitive detectors. This innovative design of a hybrid atomic-electrostatic instrument is considered as one of the promising accelerometer candidates for future space missions \citep{abrykosov_impact_2019,zahzam_hybrid_2022,hosseiniarani_advances_2024}. Our experimental setup allows us to rotate the whole instrument with rotation rates in the range of a Nadir-pointing satellite (of the order of \SI{1}{\milli \radian \per \s}) for which significant detrimental impacts on a ground cold-atom gravimeter occur despite smaller interrogation times. As another original aspect of the work, we also considered non-constant rotation velocities leading to strong Euler acceleration whose effects can also be mitigated by counter-rotating the interferometer mirror. This work aims at a deeper understanding of the various effects that limit the performance of a cold-atom accelerometer and proposes an original technical approach based on electrostatic actuation of an EA's proof-mass, coupled to the cold-atom interferometer, rather than piezoelectric actuation of a mirror mount as commonly proposed.

In this paper, we first present in Sec.~\ref{sec:level_1} the model used to derive the impact of rotations on a cold-atom accelerometer. This model considers the whole sensor rotation, the mirror rotation and their effects on both the phase shift and contrast of a Mach-Zehnder cold-atom interferometer. Then, the laboratory prototype is described, as well as the measurement protocol in Sec.~\ref{sec:level_2} and Sec.~\ref{sec:level_3}. Lastly, in Sec.~\ref{sec:level_4}, experimental contrasts and phase shifts of a rotating interferometer are investigated in the light of the model developed in the first section. The impacts of the sensor and the mirror rotations on both the phase shift and contrast are studied separately. The rotation compensation method is then characterized by considering two different angular motions: the first with an important angular velocity of \SI{1.43}{\milli \radian \per \s}  and the second with an important angular acceleration of \SI{43.5}{\milli \radian \per \square \s}.

%------------------------------------------------------------------------------
\section{\label{sec:level_1}Phase and contrast model}

\subsection{\label{sec:level_1A}Atom interferometer without rotation}
\begin{figure}[h!]
\includegraphics[scale=1]{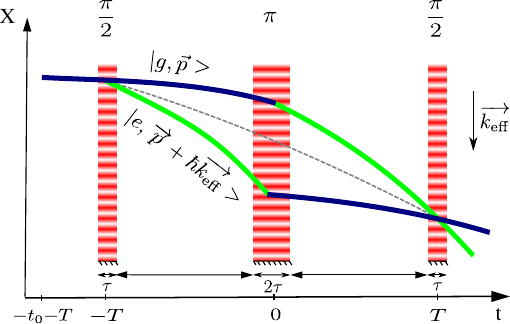}
\caption{Space-time diagram of a Mach-Zehnder atomic accelerometer measuring along the X axis. The three long red rectangles are the laser pulses $\frac{\pi}{2}-\pi-\frac{\pi}{2}$ driving the atomic transitions between the ground state $\ket{g,\vec{p}}$ and the excited state $\ket{e,\vec{p}+\hbar \vec{k}_{\mathrm{eff}}}$. The atoms free fall for a time $t_0$ before the start of the interferometer. The first and third laser pulses put the atoms in a superposition of states and have a duration $\tau$. The second pulse redirects the atom cloud and lasts $2\tau$. The free evolution of the atoms between the pulses lasts $T\gg \tau$. Experimentally, $t_0=\SI{10}{\milli\second}$, $T=\SI{46}{\milli\second}$ and $\tau=\SI{4}{\micro \second}$.}
\label{fig_1_space_time_inteferometer}
\end{figure}

Atom interferometers are based on the principle of superposition between two atomic states. Interactions between a light field and the atom drive the manipulation of quantum superposition of atomic states. In this work, we consider stimulated two-photon transitions, referred as Raman transitions, to implement atomic mirror and beam splitters \citep{kasevich_measurement_1992,borde_atomic_1989}. The studied atom interferometer is based on a Mach-Zehnder configuration \citep{kasevich_measurement_1992,peters_high-precision_2001}, involving a sequence of three Raman pulses of effective wavevector $\vec{k}_{\mathrm{eff}}$ as described in Fig.~\ref{fig_1_space_time_inteferometer}. A two-frequency laser (corresponding to the wavevectors $\vec{k}_1$ and $\vec{k}_2$), retro-reflected on a mirror, addresses Raman transitions between two stable hyperfine states of $^{87}$Rb atoms. The two states are a ground state $\ket{g,\vec{p}}$ and an excited state $\ket{e,\vec{p}+\hbar \vec{k}_{\mathrm{eff}}}$ with $\vec{p}$ the initial momentum of the atom. After a cooling step, the atoms are released in free fall at time $t=-t_0-T$. The atoms fall for a time $t_0$ before the beginning of the interferometer, defined as the time of the first laser pulse. The total interrogation time of the interferometer is $2T$, $T$ being the time separation between two consecutive laser pulses.\\

The laser effective wavevector $\vec{k}_{\mathrm{eff}}$ is the difference between one wavevector of the incoming laser $\vec{k}_{1}$ and one of the reflected laser $\vec{k}_{2}$.

\begin{equation}\label{eq_wavevector}
 \vec{k}_{\mathrm{eff}}(t)=\vec{k}_1(t)-\vec{k}_2(t). 
\end{equation}

For a single atom, the interferometer output phase shift $\Delta \Phi$ is a linear combination of the laser phases $\varphi(t)$ at the three interaction times \citep{peters_high-precision_2001} as expressed in Eq.~(\ref{eq_phase_general}). In this calculation, the laser pulses are considered infinitely short and the mirror and beamsplitter laser pulses are considered perfect. $\Delta \Phi$ is thus expressed as follows:

\begin{equation}\label{eq_phase_general}
 \Delta \Phi = \varphi\left(-T\right)-2\varphi(0)+\varphi(T).
\end{equation}
where $\varphi(t)=\vec{k}_{\mathrm{eff}}(t) \cdot \vec{r}(t) $ is the laser phase at time $t$ for a plane wave with $\vec{r}(t)$ the distance vector between the retro-reflection mirror and the atom. The atom trajectory is thus approximated by the mean trajectory between the two arms of the interferometer. The mirror can be considered as the laser phase reference. The measurement axis corresponds to the direction of the effective wavevector $\vec{k}_{\mathrm{eff}}$. The instrument aims to measure the atom acceleration along $\vec{k}_{\mathrm{eff}}$ derived from the interferometer output phase shift. As a first approximation, the interferometer phase is proportional to the atom acceleration $a_{M}$ in the mirror frame. $\Delta \Phi$ can thus be approximated as follows:

\begin{equation}\label{eq_phase_approx}
\Delta \Phi \approx \vec{k}_{\mathrm{eff}} \cdot \vec{a}_{M} T^2.
%\Delta \Phi \approx - k_{\mathrm{eff}} T^2 a_{xM}
\end{equation}

In principle, the detected signal of the interferometer is the probability $P_e$ for an atom to be detected in the excited state at the output of the interferometer. It depends sinusoidally on the phase shift expressed in Eq.~(\ref{eq_phase_general}):

\begin{equation}\label{eq_Pe_1atom}
P_e = P_0-\frac{C}{2}\cos(\Delta \Phi).
\end{equation}
with $C$ the contrast and $P_0$ the offset of the interferometer output signal for one atom.\\

Experimentally, this quantity is derived from fluorescence measurements of the atomic population at the two outputs of the interferometer. The detected signal gathers the fluorescence emitted by the whole atomic cloud and therefore the measured probability $\overline{P}_e$ can be considered as the average probability over the atom cloud neglecting the detection system response:

\begin{equation}\label{eq_P_e_mean_1}
\overline{P}_e = \overline{P}_0-\frac{\overline{C}}{2}\cos(\overline{\Delta \Phi}).
\end{equation}
with $\overline{\Delta \Phi}$ the mean phase shift, $\overline{C}$ the mean contrast and $\overline{P}_0$ the mean offset of the interferometer output signal considering the whole detected atom cloud.  Note that we neglect here the response of the detection system that could be taken into account as done in \cite{farah_effective_2014}.  \\

Since each atomic phase depends on position and velocity, the average probability can be written as a function of $D_{\vec{v}}$, the velocity distribution, and $D_{\vec{r}}$, the position distribution of the atom cloud:

\begin{equation}\label{eq_P_e_mean_2}
\overline{P}_e=\iint P_e(\vec{r},\vec{v}) \, D_{\vec{v}} \, D_{\vec{r}} \: d\vec{v} \, d\vec{r}.
\end{equation}

Thus, the normalized contrast can be deduced from the phase shift variance over the atomic cloud which can be computed knowing the position and velocity distributions under the hypothesis of Gaussian velocity and position distributions:

\begin{equation}\label{eq_C_mean}
\frac{\overline{C}}{C}=\exp(- Var(\Delta \Phi)/2)
\end{equation}

Eq.~(\ref{eq_C_mean}) is obtained considering the contribution of each atom as equivalent; i.e. $C$ does not depend on the velocity and position of the atom. A more complete model could be ultimately elaborated considering this aspect \citep{gillot_limits_2016}.

%----------------------------------------------------------------------------------------------------
\subsection{\label{sec:level_1B}Rotation of the atom interferometer} 
As previously reported, the rotation of the accelerometer sensor affects the phase shift and contrast at the output of the interferometer through Coriolis, centrifugal, and Euler acceleration terms \cite{lan_influence_2012,beaufils_rotation_2023,bongs_high-order_2006,darmagnac_de_castanet_atom_2024,zahzam_hybrid_2022,hogan_light-pulse_2007,duan_suppression_2020,zhao_extension_2021}. The phase shift being determined by averaging over the atom cloud, its alteration due to rotation is related to the mean initial velocity and position of the whole atom cloud. On the other hand, the contrast loss due to rotation is related to the size and temperature of the cloud, which could lead in some cases to a complete loss of signal and consequently to the impossibility to derive an acceleration measurement \citep{lan_influence_2012,zahzam_hybrid_2022,darmagnac_de_castanet_atom_2024}.\\

\begin{figure}[h!]
\centering
\includegraphics[scale=1]{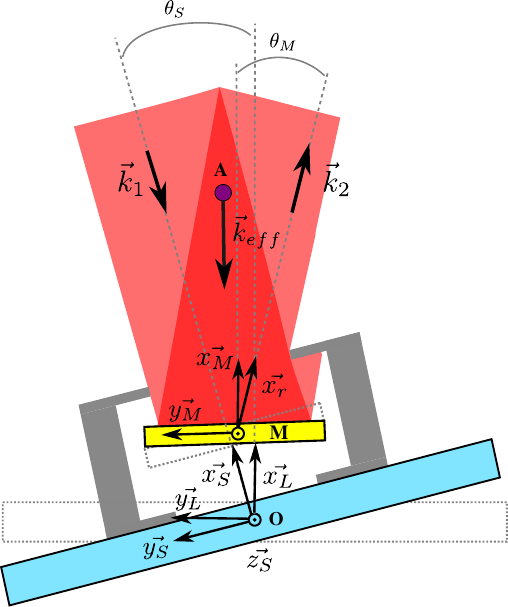} 
\caption{Description of the rotating frames: the sensor frame $\mathcal{R}_{S}$ and the mirror frame $\mathcal{R}_{M}$ in rotation in the laboratory frame $\mathcal{R}_{L}$. The blue rectangle is the table supporting the sensor. This table and the incoming laser are fixed in $\mathcal{R}_{S}$ and rotate around $O$ the center of rotation of the sensor. The yellow rectangle is the retro-reflection mirror rotating around M. $\vec{x}_r$ is the unitary vector in the direction of the reflected laser in the mirror and A the center of the atom cloud.}
\label{fig_2_frames}
\end{figure}

In this section, the calculation of the induced phase shift and contrast loss is detailed considering an atomic interferometer rotating around only one axis: $\vec{z}_S=\vec{z}_L=\vec{z}_M$. The different frames used in this calculation are illustrated in Fig.~\ref{fig_2_frames}. The directions of the unitary vectors are defined in each frame. $\mathcal{R}_{L}=\lbrace O, \vec{x}_L, \vec{y}_L, \vec{z}_L \rbrace$ is the laboratory frame, which is considered inertial in this calculation, where $O$ is the center of rotation of the sensor and $\vec{x}_L$ is the unitary vector in the opposite direction of Earth's gravity $\vec{g}$. Earth's rotation rate is considered constant during the experiment, and its effect on the interferometer can be considered negligible. $\mathcal{R}_{S}=\lbrace O, \vec{x}_S, \vec{y}_S, \vec{z}_S \rbrace$ is the sensor frame, where $\vec{x}_S$ is aligned with the direction of $\vec{k}_1$ and the normal vector to the table supporting the sensor head (in blue on Fig.~\ref{fig_2_frames}). $\mathcal{R}_{M}=\lbrace M, \vec{x}_M, \vec{y}_M, \vec{z}_M \rbrace$ is the mirror frame, where M is the center of rotation of the mirror, $\vec{x}_M$ is the normal vector to the mirror (in yellow in Fig.~\ref{fig_2_frames}), and $\vec{x}_r$ is the unitary vector in the direction of the reflected laser $\vec{k}_2$. In this section, the mirror is not rotating in the sensor frame, then $\mathcal{R}_{M} \equiv \mathcal{R}_{S}$ and $\vec{x}_r=\vec{x}_S$: $\vec{k}_1$ and $\vec{k}_2$ are perfectly counter propagating\\

The values of angle $\theta_S$ between $\vec{x}_S$ and $\vec{x}_L$ at pulse times $-T$, $0$, and $T$ can be expressed as follows: 
\begin{eqnarray}\label{eq_angle_sensor}
\theta_S(-T)       &=& \theta_S^0 - \Omega^0_S T + \frac{1}{2}\dot{\Omega}^0_S T^2,\\
\theta_S(0)      &=& \theta_S^0,\nonumber\\
\theta_S(T)  &=& \theta_S^0 + \Omega^0_S T + \frac{1}{2}\dot{\Omega}^0_S T^2,\nonumber
\end{eqnarray}

where $\Omega^0_S$ is the mean angular velocity between the third and the first laser pulse and $\dot{\Omega}^0_S$ is the mean angular acceleration over the three laser pulses sequence. 

To compute the laser phase at each pulse, the effective wavevector of the laser has to be expressed as a function of the rotation parameters. As defined in Eq.~(\ref{eq_wavevector}), the effective wavevector is the difference between the wavevector of the incoming laser $\vec{k}_{1}=-k_1 \vec{x}_S $ and the reflected laser $\vec{k}_{2}=k_2 \vec{x}_r$. The unitary vector $\vec{x}_S$ can be written as $\vec{x}_S=\cos \theta_S \vec{x}_L + \sin \theta_S \vec{y}_L$. In presence of rotation of the whole sensor, the expression of $\vec{k}_{\mathrm{eff}}$ in the laboratory frame, $\mathcal{R}_{L}$, thus reads 

\begin{equation}\label{eq_keff_sensor_rotation}
\vec{k}_{\mathrm{eff}}(t)=-{k}_{\mathrm{eff}}\begin{pmatrix}
\cos{ \theta_S(t) }\\ 
\sin{ \theta_S(t) }\\ 
0
\end{pmatrix} \simeq -{k}_{\mathrm{eff}}\begin{pmatrix}
1 - \frac{\theta_S(t)^2}{2} \\ 
\theta_S(t) \\ 
0
\end{pmatrix} .
\end{equation}
with ${k}_{\mathrm{eff}}=k_1+k_2$ the norm of $\vec{k}_{\mathrm{eff}}$ in the absence of rotation. The direction of the wavevector is impacted by the sensor rotation but not its norm.

We can decompose the classical trajectory of the mean position of the atomic wave packet $\vec{r}=\overrightarrow{MA}=\overrightarrow{MO}+\overrightarrow{OA}$ as a function of the position of the atom cloud relative to the sensor center of rotation $(x_{OA},y_{OA},z_{OA})$, the relative position of the mirror to the sensor center of rotation $(x_{OM},y_{OM},z_{OM})$, the velocities $(v_{x},v_{y},v_{z})$ and accelerations $(a_{x},a_{y},a_{z})$ of the atomic cloud at the first interferometric pulse in a second lab frame considered inertial which coincide with the sensor frame at the first pulse $\mathcal{R}_{L}'=\mathcal{R}_{S}(t=-T)$. Expressing the atomic trajectory in this frame simplifies the phase shift and contrast expressions. Last, this choice explains the seeming differences between the equations of this work and previous literature. We finally obtained:\\
\begin{equation}\label{eq_atom_trajectory}
\vec{r}(t)=\begin{pmatrix}
a_{x}\frac{(t+T)^2}{2} & + & v_{x}(t+T) & + & x_{MA}\\ 
a_{y}\frac{(t+T)^2}{2} & + & v_{y}(t+T) & + & y_{MA}\\
a_{z}\frac{(t+T)^2}{2} & + & v_{z}(t+T) & + & z_{MA}\\
\end{pmatrix}.
\end{equation}
the mean atom trajectory in the second lab frame $\mathcal{R}_{L}'$ at the first laser pulse.
The different parameters of the atomic cloud trajectory at the first pulse have to be computed as a function of the kinetic parameters at the launch time $t=-t_0-T$ given in Tab.~\ref{tab_cloud_parameters}. We consider a non zero initial velocity $\vec{v}_{res}$ in the lab frame $\mathcal{R}_{L}$ that can result from imbalanced laser beams during the molasses stage \citep{farah_effective_2014,werner_laser_1993}. The sensor rotation impacts the magneto-optical trap (MOT) position and velocity at the atoms release and has to be taken into account through the MOT angular position $\theta_{MOT}$ and instantaneous angular velocity $\vec{\Omega}_{MOT}$. The atomic velocity at launch is then $\vec{v}_0=\vec{v}_{res}+\vec{\Omega}_{MOT} \wedge \vec{r}_0$ with $\vec{r}_0=\overrightarrow{OA}(-t_0-T)$. The only considered acceleration, $\vec{a}_0=a_{x_0} \vec{x}_L$ takes Earth's gravity into account (the Earth-gravity gradient is neglected). The velocity in $\mathcal{R}_{L}'$ is then: $\vec{v}_1=\vec{v}_0+\vec{a}_1 t_0+0.5 \vec{v}_{rec}$ with $\vec{v}_{rec}$ the recoil velocity and $\vec{a}_1$ the acceleration in $\mathcal{R}_{L}'$ leading to the value of $(v_{x},v_{y},v_{z})$.\\
Using Eqs~(\ref{eq_phase_general}), (\ref{eq_keff_sensor_rotation}), and (\ref{eq_atom_trajectory}), the phase shift of the rotating interferometer can be expressed as:

\begin{eqnarray}\label{eq_phase_sensor}
&&\overline{\Delta \Phi}_S = k_{\mathrm{eff}} T^2 \Bigg[- a_{x} \\
												&&  - 2 \Omega^0_S \cdot \left(v_{y}+ \frac{3}{2} a_y T \right) \nonumber\\
                        &&  -\dot{\Omega}^0_S \cdot \left( y_{OA} + v_{y} T + \frac{1}{2} a_y T^2 \right) \nonumber\\
                        &&  +{\Omega^0_S}^2 \cdot \left( x_{OA} +3 v_{x} T + \frac{7}{2} a_{x}T^2 \right) \Bigg].\nonumber
\end{eqnarray}
Eq.~(\ref{eq_phase_sensor}) is an approximation as terms scaling in ${\Omega^0_S}^3$, ${\dot{\Omega}^0_S}^2$, $\dot{\Omega}^0_S \Omega^0_S$ and $\theta_S^0$ are neglected. The phase shift thus depends on the atoms' vertical acceleration and is impacted by the three inertial accelerations as follows:
\begin{itemize}
\item Coriolis acceleration: $ - 2 \Omega^0_S \cdot \left( v_{y}+ \frac{3}{2} a_y T \right)$;
\item Euler acceleration:
$-\dot{\Omega}^0_S \cdot \left( y_{OA} + v_{y} T + \frac{1}{2} a_y T^2 \right)$;
\item Centrifugal acceleration: \\
${\Omega^0_S}^2 \cdot \left( x_{OA} +3 v_{x} T + \frac{7}{2} a_{x} T^2 \right)$.
\end{itemize}

Using Eq.~(\ref{eq_P_e_mean_2}) and assuming independent Gaussian distributions with standard deviations $\sigma_x$, $\sigma_y$, $\sigma_z$ for the initial position and $\sigma_{v_x}$, $\sigma_{v_y}$, $\sigma_{v_z}$ for the velocity, the expression of the contrast decay as a function of the rotation parameters reads
\begin{eqnarray}\label{eq_contrast_sensor}
\frac{\overline{C}_S}{C} = \exp \Bigg[ -\frac{k_{\mathrm{eff}}^2 T^4 }{2}&& \Bigg( (2\Omega^0_S+T\dot{\Omega}^0_S)^2 \cdot \sigma_{v_y}^2\\
													&& + {\dot{\Omega}^0_S}^2 \cdot \sigma_y^2 \nonumber\\
													&& + {\Omega^0_S}^4 \cdot \left[ \sigma_x^2 + (3T)^2 \sigma_{v_x}^2\right]\Bigg) \Bigg]. \nonumber
\end{eqnarray}
%
%----------------------------------------------------------------------------------------------------
\subsection{\label{sec:level_1C} Rotation of the mirror}
In this section, we detail the impact on the interferometer output of the retro-reflection mirror rotation only. The considered mirror rotation vector, $\vec{\Omega}_M$, is collinear to $\vec{z}_M$ axis, as shown in Fig.~\ref{fig_2_frames}. Similarly to the case where the rotation is applied to the whole instrument, the angle $\theta_M$ between $\vec{x}_M$ and $\vec{x}_S$ can be expressed at times $-T$, $0$, and $T$ as a function of the mirror mean angular velocity, $\Omega^0_M$, the mirror mean angular acceleration, $\dot{\Omega}^0_M$, and the mirror angle at the second pulse $\theta_M^0$ in the same way as in Eq.~(\ref{eq_angle_sensor}):
\begin{eqnarray}\label{eq_angle_mirror}
\theta_M(-T)       &=& \theta_M^0 - \Omega^0_M T + \frac{1}{2}\dot{\Omega}^0_M T^2,\\
\theta_M(0)      &=& \theta_M^0,\nonumber\\
\theta_M(T)  &=& \theta_M^0 + \Omega^0_M T + \frac{1}{2}\dot{\Omega}^0_M  T^2, \nonumber
\end{eqnarray}

To calculate the laser phase at each pulse time, the effective wavevector is expressed as a function of the mirror and the sensor rotation parameters. The different unitary vectors can be expressed with the help of the mirror angle: $\vec{x}_M=\cos \theta_M \vec{x}_S + \sin \theta_M \vec{y}_S$. The vector in the direction of the reflected laser is computed as follows:
\begin{equation}
\vec{x}_r=- \vec{x}_S+2 (\vec{x}_M.\vec{x}_S) \vec{x}_M.
\end{equation}

As for the rotation of the whole sensor, the phase shift and contrast are in the same way impacted by the mirror rotation. The phase shift is computed using Eq.~(\ref{eq_phase_general}).
\begin{eqnarray}\label{eq_phase_mirror}
&&\overline{\Delta \Phi}_M = k_{\mathrm{eff}} T^2 \Bigg[-a_x \\
												&& -2\Omega^0_M \cdot \left( v_y + a_y T \right) \nonumber\\
												&& -\dot{\Omega}^0_M \cdot \left( y_{MA} + v_y T + a_y T^2 \right) \nonumber\\
												&& +2{\Omega^0_M}^2 \cdot \left(x_{MA} +v_x T + a_x T^2  \right]) \Bigg],\nonumber
\end{eqnarray}
with $(x_{MA},y_{MA},z_{MA})$ the atom-cloud position relative to the mirror's center of rotation at the first pulse. Eq.~(\ref{eq_phase_mirror}) is an approximation as terms scaling in ${\Omega^0_M}^3$, ${\dot{\Omega}^0_M}^2$, $\dot{\Omega}^0_M \Omega^0_M$ and $\theta_M^0$ are neglected. This equation is similar to Eq.~(\ref{eq_phase_sensor}) except for the contribution of the centrifugal acceleration which is multiplied by 2 when the mirror is rotating as the magnitude of the Raman laser wavevector now depends on $\theta_M(t)$:
\begin{equation}\label{eq_keff_mirror_rotation}
\vec{k}_{\mathrm{eff}}(t)=-\begin{pmatrix}
k_1 + k_2 \cos{ 2\theta_M(t) }\\ 
k_2\sin{ 2 \theta_M(t) }\\ 
0
\end{pmatrix} \simeq -k_{\mathrm{eff}}\begin{pmatrix}
1- \theta_M^2(t) \\ 
 \theta_M(t) \\ 
0
\end{pmatrix},
\end{equation}
The contrast decay due to the rotation of the mirror is also very similar to the contrast decay due to the rotation of the whole sensor in Eq.~(\ref{eq_contrast_sensor}). The only difference concerns once again the impact of the centrifugal acceleration:
\begin{eqnarray}\label{eq_contrast_mirror}
\frac{\overline{C}_M}{C} = \exp \Bigg[ -\frac{k_{\mathrm{eff}}^2 T^4 }{2}&& \Bigg( (2\Omega^0_M+T\dot{\Omega}^0_M)^2 \cdot \sigma_{v_y}^2\\
													&& + {\dot{\Omega}^0_M}^2 \cdot \sigma_y^2 \nonumber\\
													&& + {\Omega^0_M}^4 \cdot \left[ 4\sigma_x^2 + (2T)^2 \sigma_{v_x}^2\right]\Bigg) \Bigg]. \nonumber
\end{eqnarray}
The similarities between the impacts of the sensor rotation and the mirror rotation on interferometer output show that counter-rotating the interferometer mirror offers a possibility to mitigate rotation issues. Nevertheless, we can already see that both rotations are not completely equivalent and a basic compensation scheme will not allow to fully remove all the detrimental impact of rotation. This will be studied in the next section.
%
%----------------------------------------------------------------------------------------------------
\subsection{\label{sec:level_1D} Rotation compensation}
In order to limit the detrimental impact of rotation on the atomic accelerometer, we consider here the rotation compensation method consisting in rotating the retro-reflection mirror with an opposite angle, leaving the mirror rotation-less in the laboratory frame during the interferometric phase. Thanks to the compensation, the direction effective wavevector is constant during the interferometer but not its norm:\\
\begin{equation}\label{eq_keff_comp}
\vec{k}_{\mathrm{eff}}(t)= -{k}_{\mathrm{eff}}\begin{pmatrix}
1 - \frac{\theta_S(t)^2}{2} \\ 
0 \\ 
0
\end{pmatrix} .
\end{equation}

Let us first analyze the contrast of the interferometer in this rotation compensation scheme. Based on the analysis conducted in the previous sections, considering the combined effect of both the rotation of the sensor and the rotation of the mirror, one can derive the contrast of the atom interferometer:
\begin{equation}\label{eq_contraste_compensation}
\frac{\overline{C}^{comp}_{M=S}}{C}=\exp \Bigg[ -\frac{k_{\mathrm{eff}}^2 T^4}{2} {\Omega^0_S}^4 \cdot \left[ \sigma_x^2 + \sigma_{v_x}^2 T^2 \right] \Bigg].
\end{equation}
As shown in the above expression, the contrast is not impacted anymore by the Coriolis and Euler accelerations. The only remaining source of contrast loss is the unchanged contribution of the centrifugal acceleration of Eq.~(\ref{eq_phase_sensor}).\\

Considering the phase shift of the atom interferometer, an analogous approach yields:
\begin{eqnarray}\label{eq_phase_compensation_general}
&&\overline{\Delta \Phi}_{M+S} = k_{\mathrm{eff}} T^2 \Bigg[-a_{x} \\
												&& -2(\Omega^0_M + \Omega^0_S) \cdot \left (v_{y} + a_y T \right) -\Omega^0_S a_y T \nonumber\\
												&& -(\dot{\Omega}^0_S + \dot{\Omega}^0_M) \cdot \left(y_{OA} + v_y T + \frac{1}{2} a_y T^2 \right) \nonumber \\
                                                &&-\dot{\Omega}^0_M \cdot \left( y_{MO} + \frac{1}{2} a_y T^2 \right)\nonumber\\
												&& +2{\Omega^0_M}^2 \cdot(x_{OA}+x_{MO}+T v_x + a_x T^2)\nonumber\\
                                                && + {\Omega^0_S}^2 \cdot(x_{OA}+3T v_x + \frac{7}{2} a_x T^2)\nonumber\\
                                                && + 2\Omega^0_M \Omega^0_S \cdot(x_{OA}+x_{MO}+2T v_x + 2a_x T^2) \Bigg]. \nonumber
\end{eqnarray}
If Eq.~(\ref{eq_phase_compensation_general}) seems different from the ones found in previous articles such as \citep{beaufils_rotation_2023}, one can easily retrieve the expressions from the literature in a simple case : no time delay between the atoms launch and the beginning of the interferometer $t_0=0$ and null angular accelerations $\dot{\Omega}^0_S=\dot{\Omega}^0_M=0$. Then, the atoms mean velocity can be expressed as  $\vec{v}_1=\vec{v}_0+\vec{\Omega}_S\wedge \overrightarrow{OA}$ and then $v_{y1}=v_{y0}+\Omega_S^0 x_{OA}$ with $\vec{v}_0$ the velocity at the end of the cooling stage. Finally, the Coriolis term becomes:\\
\begin{eqnarray}
- 2 v_{y1} \cdot (\Omega_S^0+\Omega_M^0)= && - 2 v_{y0} \cdot (\Omega_S^0+\Omega_M^0)\\
&&- 2 \Omega_M^0 \Omega_S^0 \cdot x_{OA}-2{\Omega_S^0}^2 \cdot x_{OA}. \nonumber
\end{eqnarray}
leading to a similar expression as in \citep{beaufils_rotation_2023}.

The induced phase shift of Eq.~(\ref{eq_phase_compensation_general}) thus depends on both center-of-rotation positions $x_{MO}$ and $y_{MO}$, which limit the effect of the rotation compensation method. In a ideal rotation compensation scenario, the sensor and the mirror rotate in exactly opposite ways, each around its own center of rotation. Their angular positions will be exactly opposite $\theta_{M}(t)=-\theta_{S}(t)$, at least at the three laser pulses times, as will be the mean angular velocities and accelerations: $\Omega^0_M=-\Omega^0_S$ and $\dot{\Omega}^0_M=-\dot{\Omega}^0_S$. Some terms in the general expression of the phase shift, in Eq.~(\ref{eq_phase_compensation_general}), are thus canceled out by the rotation compensation, leading to:
\begin{fleqn}
\begin{eqnarray}\label{eq_phase_compensation}
\overline{\Delta \Phi}^{comp}_{M=S} = && k_{\mathrm{eff}} T^2 \Bigg[-a_{x} \\
                                        && -\Omega^0_S \cdot a_y T \nonumber\\
										&&-\dot{\Omega}^0_M \cdot \left( y_{MO} + \frac{1}{2} a_y T^2 \right)\nonumber\\
										&& +{\Omega^0_S}^2 \cdot \left[\right.  x_{OA} + 2x_{MO} + v_{x} T + \frac{3}{2} a_{x} T^2   \left] \right. \Bigg].\nonumber
\end{eqnarray}
\end{fleqn}

In these conditions, the rotation induced phase shift is reduced. The phase shift is not impacted anymore by the Coriolis acceleration in the absence of a transverse acceleration $a_y$. However, a Euler acceleration term is still present due to a potential transverse misalignment $y_{MO}$ between the mirror and the sensor centers of rotation. In principle, this Euler acceleration term could be canceled out with a precise relative adjustment of the atom-cloud position and the sensor center of rotation, neglecting any effects impacting atom-cloud position stability.\\
Moreover, several centrifugal acceleration terms remain. One of them is due to the potential distance between the mirror and the sensor centers of rotation, leading to a phase shift scaling as $x_{MO}$. The rest of the centrifugal acceleration term: $ x_{OA} + x_{MO} + v_{x} T + \frac{3}{2} a_{x} T^2$  is explained by the variation of the wavevector magnitude due to the mirror rotation (Eq.~\ref{eq_keff_comp}). Note that this result was already highlighted recently in \citep{beaufils_rotation_2023,darmagnac_de_castanet_atom_2024}. To reduce this contribution, \citep{beaufils_rotation_2023} proposed a configuration that involved both the rotation of the incident Raman beam and the mirror in the case of an atom accelerometer onboard an Earth orbiting satellite.
Regarding the centrifugal acceleration terms, the existing design of our instrument doesn't allow the cancellation of all the related contributions.\\
However, we can propose some potential alternative designs. A first possibility would be to apply two frequency jumps on the Raman laser frequency before the second and before the third laser pulse, compensating the effect of wavevector magnitude variation during the mirror rotation. A similar technique was already proposed and implemented on the central $\pi$ pulse to compensate for the gravity gradient phase \citep{roura_circumventing_2017, Damico_Cancel_2017, Overstreet_Eff_2018, Caldani_Simu_2019} . This should allow, jointly with matching the mirror and sensor centers of rotation, to remove the centrifugal acceleration terms. Considering, for instance, a constant angular velocity of typically $\Omega_S = 1.1$ mrad/s, representative of an Earth orbiting satellite, the variation of the effective wavevector to compensate between the first and second pulses is $\Delta ||\vec{k}_{\mathrm{eff}}||= \frac{(\Omega_S T)^2}{2} ||\vec{k}_{\mathrm{eff}}||$. This would lead to a first frequency jump of $\approx \SI{465}{\mega \hertz}$ and a second one of $\approx \SI{1396}{\mega \hertz}$ if we consider an interrogation time $T=1$ s.\\
A second possibility would be, as can be seen in Eq.~(\ref{eq_phase_compensation}), to configure the position of the atom cloud $A$, the mirror $M$ and the center of rotation of the platform $O$ so to cancel the term $x_{OA} + 2x_{MO}$, corresponding to a mirror in between, at the same distance from each other. Such a solution would cancel the centrifugal contribution assuming the terms in Eq.~(\ref{eq_phase_compensation}) scaling with the acceleration $a_{x}$ and velocity $v_{x}$, remain negligible. These propositions are very preliminary and should be in any case investigated further in detail to assess their potential of interest for future atom accelerometer application involving detrimental rotation impacts, especially in the case of an Earth orbiting satellite in Nadir pointing mode. These considerations come for instance without considering the impact of gravity gradients.\\
The rotation compensation method described from a theoretical point of view in this section is also implemented experimentally in the following. The experimental setup is described in the next section.

\section{\label{sec:level_2}Experimental setup}
\begin{figure}[h!]
\centering
\includegraphics[scale=1]{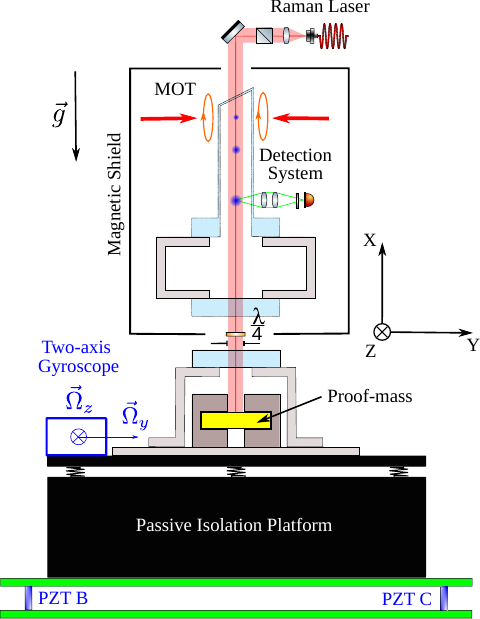}
\caption{Experimental setup composed of a cold atom gravimeter, an electrostatic accelerometer, a two axis gyroscope, a passive isolation platform and piezo-electric actuators (PZT B and PZT C) allowing the whole setup to be rotated.} 
\label{fig_3_setup}
\end{figure}

Here, we present the architecture of the experimental setup used to analyze the impact of rotation on the output of a cold-atom accelerometer. As depicted in Fig.~\ref{fig_3_setup}, it is constituted of a cold-atom gravimeter standing above an electrostatic accelerometer whose proof-mass is employed as a retro-reflection mirror for the atom interferometer laser. The whole experiment and a two-axis gyroscope are installed on a passive vibration isolation platform. This platform is necessary to mitigate the impact of ground vibrations on the interferometer. The isolation platform is mounted on a table that is supported by piezoelectric actuators. This configuration allows the rotation of the sensor head around the $Z$ axis by driving the height of the piezoelectric actuator B. The sensor rotation $\Omega_S(t)$ is measured by the two-axis gyroscope.\\
The atom interferometer is similar to the one described in \cite{bidel_compact_2013}. The laser frequencies necessary to the experiment are generated by phase modulation of a frequency-doubled telecom laser system \cite{carraz_compact_2009}. Upstream the interferometer, a 3D magneto-optical trap of rubidium 87 is loaded from a background vapor. After a stage of optical molasses and microwave selection, an atom cloud with a typical temperature of \SI{2}{\micro\kelvin}  is obtained in the ground state $\ket{5²S_{1/2},F=1,m_F=0}$ \citep{bidel_compact_2013}. After a free fall of $t_0=\SI{10}{\milli\second}$, a vertical Mach-Zehnder atom interferometer is completed. Three equally spaced laser pulses realize two-photon Raman transitions between the hyperfine states $\ket{g}=\ket{5²S_{1/2},F=1,m_F=0}$ and $\ket{e}=\ket{5²S_{1/2},F=2,m_F=0}$. The first and third laser pulses last $\tau=\SI{4}{\micro \second}$ and the second one lasts $2\tau=\SI{8}{\micro \second}$, to respectively deliver $\pi/2$ and $\pi$ pulses. The time delay between the consecutive pulses is $T=\SI{46}{\milli\second}$ characterizing the interferometer interrogation time. The two frequencies necessary to the two-photon Raman transition are generated by retro-reflecting a phase modulated laser on a mirror  \citep{carraz_compact_2009}.

\begin{figure}[h!]
\centering
\includegraphics[width=0.3\textwidth]{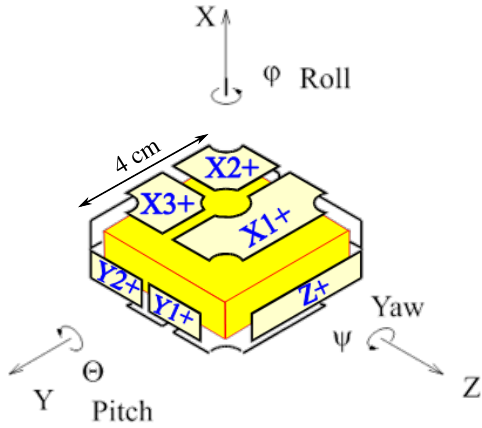} 
\caption{Electrostatic accelerometer proof-mass and electrodes. The arrangement of the electrodes allows the control of the six degrees of freedom of the proof-mass.}
\label{fig_4_EA}
\end{figure}

As already mentioned, the retro-reflection mirror for the cold-atom interferometer is the proof-mass of an electrostatic accelerometer (EA). This configuration allows for the hybridization of both sensors based on different technologies  \cite{merlet_operating_2009,geiger_detecting_2011,lautier_hybridizing_2014,bidel_absolute_2018,zahzam_hybrid_2022}. For more details on the design and operation of electrostatic accelerometers, one can, for example, refer to \cite{rodrigues_space_2022}. In contrast to a standard EA, a glass window has been added to the EA's vacuum chamber to allow the laser to be reflected on the EA proof-mass. The core of the EA is made of gold-coated (Ultra Low Expansion glass) with a side of 4 cm and a mass of 35 g in electrostatic levitation at the center of a cage holding all the electrodes (see Fig.~\ref{fig_4_EA}). The position of the proof-mass can be measured along the six degrees of freedom owing to a differential capacitance detection. The position of the proof-mass is controlled by use of a servo feedback control on the applied electrostatic forces. In the acceleration measurement mode, the acceleration value is deduced from the electrostatic force applied to maintain the proof-mass at the center of the cage. Here, the acceleration measurement of the EA is not used since the EA is operated only as a precise actuated mirror and the position of the proof-mass according to the six-degree of freedom is driven by the electrostatic forces.\\
\section{\label{sec:level_3}Experimental methods}
\subsection{\label{sec:level_3A}Atom interferometer rotation}
\begin{figure}[h!]
\centering
\includegraphics[width=0.45\textwidth]{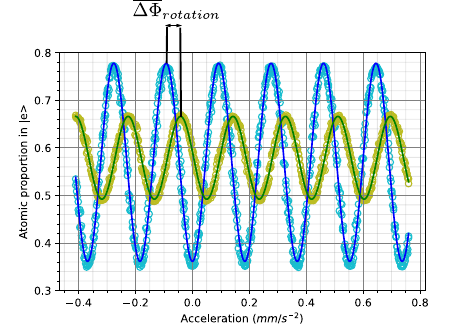} 
\caption{Atomic interference fringes collected by scanning the two-photon Raman laser frequency rate $\alpha$. The horizontal axis is the pseudo acceleration induced by the scan of $\alpha$. The blue, respectively green, circles are the measured proportion of atoms in the excited state at the output of the interferometer in the absence, respectively in the presence, of a mirror rotation. Solid blue, respectively green, line is the sinusoidal fit of the experimental data in the absence, respectively in the presence, of a mirror rotation with a mean angular acceleration is $\dot{\Omega}^0_M=\SI{-52.4}{\radian \per \square \second}$ and a mean angular velocity close to zero.}
\label{fig_5_fringes}
\end{figure}

To study the impact of rotations, we need to measure the contrast and phase shift of the interferometer. During the interferometer phase, the two-photon Raman laser frequency is chirped with a rate $\alpha \approx 25$ MHz/s to compensate for the Doppler shift of the free-falling atoms. The scan of the frequency chirp rate $\alpha$ allows us to sweep artificially the acceleration seen by the atoms. This method enables us to obtain atomic interference fringes from which the phase shift and contrast of the interferometer can be deduced (see Fig.~\ref{fig_5_fringes}). Typically, the fringes display a contrast without rotation of $C\approx0.42$ and are obtained after scanning over 400 measurements points, each corresponding to an experimental cycle.

To rotate the whole sensor, a time-dependent voltage is applied to the piezoelectric actuator B (see Fig.~\ref{fig_3_setup}). To benefit from a smooth dynamic of the sensor and thus not excite vibrations mode of the instrument mechanical structure, we choose to apply a sinusoidal input voltage at a \SI{4}{\hertz} frequency on the piezoactuator, which is the repetition rate of the experimental cycle. In this way, the piezoactuator input voltage is synchronized with the interferometer measurement cycle. The vibration isolation platform helps isolating the setup from the ground vibrations, but also from the high-frequency vibrations generated by the piezoelectric actuator. The setup generates a rotation along the $\vec{z}_S$ axis with an angular range of the order of \SI{100}{\micro \radian} leading to a mean angular velocity around \SI{2}{\milli \radian \per \s} and an acceleration of \SI{60}{\milli \radian \per \square \s} measured by the gyroscope.

As for the retro-reflection mirror, it can be rotated or translated by driving the electrostatic force applied between the electrodes and the proof-mass. The angle between the proof-mass and the electrodes (see Fig.~\ref{fig_4_EA}) is measured by capacitive detection of the electrostatic accelerometer.
\subsection{\label{sec:level_3B}Removal of systematic side effects}
To process the experimental results and focus on the effect of the rotation compensation method, which is the main purpose of this paper, two side phenomena are corrected from the experimental data.

Firstly, as the electrostatic proof-mass is rotated, small residual accelerations or rotations of the mirror are present as a result of the asymmetric electrodes architecture. For example, for an imposed angular ramp of angular velocity \SI{2}{\milli \radian \per \s} along the $\vec{z}_M$ axis, a vertical acceleration of \SI{-0.36}{\micro \meter \per \s} is measured as well as a rotation along $\vec{y}_M$ with a mean angular velocity of \SI{29}{\micro \radian \per \s} and a mean angular acceleration of \SI{-20}{\micro \radian \per \square \s}. These parasitic movements are measured by capacitive detection and the induced phase shift is estimated by use of the capacitive detection measurements. The method is described in the Supplementary Materials.

Secondly, although the rotation of the whole sensor occurs mainly along the $\vec{z}_S$ axis according to the applied \SI{4}{\hertz} sinusoidal signal, there is a small unwanted residual rotation along the $\vec{y}_S$ axis with an amplitude reaching $\approx$ 8\% of the main rotation along the $\vec{x}_S$ axis and a phase shift of $\approx$ \SI{0.5}{\radian} leading to a maximal mean angular velocity of \SI{0.13}{\milli \radian \per \s} and an acceleration of \SI{3.9}{\milli \radian \per \square \s}. The impact of this parasitic rotation on the phase shift can be computed as described in the Supplementary Materials. 
\subsection{\label{sec:level_3C}Kinematic parameters of the cloud}
To analyze the effect of the rotation, the kinematic parameters of the atom cloud such as its temperature, size, mean velocity, and mean position relative to the center of rotation of the mirror need to be measured. These parameters were determined by rotating the retro-reflection mirror as described in the Supplementary materials. Table \ref{tab_cloud_parameters} summarizes the parameters in the laboratory frame used to estimate the phase shift and contrast.

\begin{table}[h!]
\centering
\caption{Kinematic parameters of the atom cloud at launch.}
\label{tab_cloud_parameters}
\begin{tabular}{ | c | c |  c | } 
  \hline
  Parameter & Value & Type of measurement\\ 
  \hline
 $v_{x_0}$  & \SI{0(3)}{\mm \per \second} & from \citep{bidel_absolute_2020}\\ 
  \hline
 $v_{y_0}$  & \SI{-1.3(0.3)}{\mm \per \second} &  Mirror rotation   \\ 
  \hline
 $v_{z_0}$  & \SI{0.3(0.3)}{\mm \per \second} &  Mirror rotation   \\ 
  \hline
 $\sigma_{v_x}$  & \SI{11.3(0.2)}{\mm \per \second} & Raman spectroscopy    \\ 
  \hline
 $\sigma_{v_y}$  & \SI{10.8(0.2)}{\mm \per \second} &   Mirror rotation  \\ 
  \hline
 $\sigma_{v_z}$  & \SI{11.1(0.2)}{\mm \per \second} &   Mirror rotation  \\ 
  \hline
 $x_{MA}$  & \SI{420(5)}{\mm} &  By construction   \\ 
  \hline 
 $y_{MA}$  & \SI{1.09(0.03)}{\mm} & Mirror rotation   \\ 
  \hline 
 $z_{MA}$  & \SI{0.66(0.03)}{\mm} &   Mirror rotation   \\ 
  \hline 
 $\sigma_x$   & \SI{0.5}{\mm} &  from \citep{bidel_absolute_2020} \\ 
  \hline 
 $\sigma_y$   & \SI{0.42(0.05)}{\mm} &  Mirror rotation   \\ 
  \hline 
 $\sigma_z$   & \SI{0.66(0.05)}{\mm} &  Mirror rotation   \\ 
   \hline
 $x_{OA}$  & \SI{550(5)}{\mm} &  By construction   \\ 
  \hline 
 $y_{OA}$  & $\in \left[ -5.7 ; \SI{-5}{\mm} \right]$ & Fit parameter \\ 
 %-5,7 -5 -5.1 
  \hline 
 $z_{OA}$  & $\in \left[ 12.2 ; \SI{10.6}{\mm} \right]$ & Fit parameter \\  

  \hline 
\end{tabular}
\end{table}
For example, the initial mean velocity along the $\vec{y}_L$ axis, can be non zero due to intensity imbalance of the laser beams during the cooling stage \cite{werner_laser_1993,farah_effective_2014}. In our experiment, this velocity is estimated at $v_{y_0}=\SI{-1.3}{\milli \meter \per \second}$. The distance along the $\vec{x}_L$ axis between the atom cloud and the center of rotation of the sensor, located on the upper table of the isolation platform, is estimated by construction, whereas the same distance along the horizontal plane is a free parameter chosen to fit the experimental data within a plausibility range of a few centimeters.
\section{\label{sec:level_4}Experimental results}
\subsection{\label{sec:level_4A}Rotation of the atom interferometer}
An experimental study to investigate the impact of rotation on the interferometer output is conducted by rotating the whole setup at 4 Hz following the procedure described in Sec.~\ref{sec:level_3}.

\begin{figure}[h!]
\centering
\includegraphics[scale=1]{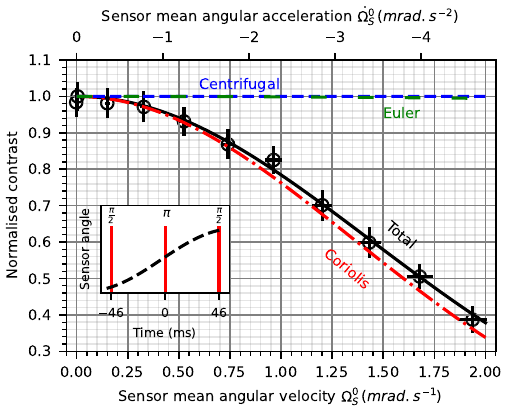} 
\caption{Impact of the sensor rotation on the interferometer contrast. The black circles represent the experimental data with the $1 \sigma$ error bars. The black line is the theoretical contrast of Eq.~(\ref{eq_contrast_sensor}). In the inset, the black dashed line is the sensor sinusoidal angular movement $\theta_S(t)$ deduced from the gyroscope measurement. The vertical red lines represent the temporal position of the atom interferometer laser pulses.}
\label{fig_6_Contrast_Sensor_lin}
\end{figure}

\begin{figure}
\centering
\includegraphics[scale=1]{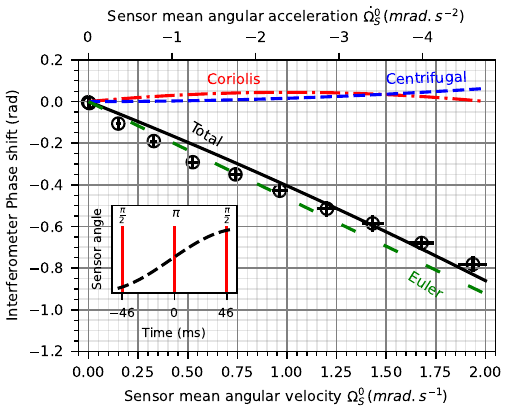} 
\caption{Impact of the sensor rotation on the interferometer phase shift. The black line is the phase shift as expected by Eq.~(\ref{eq_phase_sensor}). The experimental data are corrected from the side effects (see Section \ref{sec:level_3B}). Legend similar to Fig.~\ref{fig_6_Contrast_Sensor_lin}.}
\label{fig_7_Phase_Sensor_lin}
\end{figure}

The results in terms of contrast loss are reported in Fig.~\ref{fig_6_Contrast_Sensor_lin}. The error bars associated to the experimental data come from the sinusoidal fit of the gyroscope signal and the uncertainty of the gyroscope measurement for the uncertainty on the sensor mean angular velocity. The uncertainty on the normalized contrast is due to the sinusoidal fit errors of the interference fringes and the contrast variation over time due to experimental limits. The main source of contrast loss is due to the effect of Coriolis acceleration linked to the velocity dispersion of the atom cloud and, more specifically, to the term scaling as $4{\Omega^0_S}^2 \sigma_{v_y}^2$, in the red dashed dotted line in Fig.~\ref{fig_6_Contrast_Sensor_lin}. While the centrifugal acceleration does not contribute significantly to the loss of contrast, the Euler acceleration plays a larger role taken into account to fit the model to the experimental data. If the total model (black line) agrees with the experimental data, the Coriolis contribution is not the only contribution to the contrast loss. For this data set, the angular acceleration is non-zero due to the imperfect synchronization between the 4-Hz atomic interferometry cycle and the 4-Hz sinusoidal excitation of the sensor rotation platform. If the mean angular acceleration had been null $\dot{\Omega}_S=0$, the expected contrast loss would have been lower (red line). Notably, the contribution due to Euler acceleration has an opposite sign compared to the one due to Coriolis acceleration, so the contrast is higher in the presence of an angular acceleration, as predicted by Eq.~(\ref{eq_contrast_sensor}). This angular movement, where the second Raman laser pulse corresponds approximately to the midpoint of the sinusoid, should have cancelled the Euler acceleration. Nevertheless, a small imperfection of the angular movement phase relative to the temporal position of the laser pulses gives rise to an impact of Euler acceleration on the output of the interferometer.\\
In addition to the contrast loss, a phase shift induced by the rotation is observed (Fig.~\ref{fig_7_Phase_Sensor_lin}). The phase shift uncertainty is computed from the sinusoidal fit errors of the interference fringes and the phase shift fluctuations over time. The phase shift is significantly impacted by all the inertial accelerations and quite well reproduced by the model based on Eq.~(\ref{eq_phase_sensor}) using the gyroscope data with an error to the model below \SI{0.05}{\radian}. This error could be explained by variations of the distance $y_{OM}$ due to imperfection of the rotation actuation. The Coriolis contribution is linked to the atomic cloud transverse velocity due to by both the residual velocity at the end of the cooling stage and the MOT rotation at launch. The last depends of the sinusoidal angular movement explaining the quadratic impact of the Coriolis acceleration. The centrifugal acceleration has a quadratic impact on the phase shift and is also due to the vertical distance between the atom cloud and the sensor center of rotation. The Euler acceleration has a linear impact on the phase shift and is explained by the transverse distance between the atom cloud and the center of rotation. In this configuration, which is supposed to lower the impact of the angular acceleration, we can nonetheless observe a dominant Euler acceleration, reaching nearly \SI{1}{\radian} of atom interferometer phase shift.
\subsection{\label{sec:level_4B}Rotation of the mirror}
In this section, we present the experimental results obtained by rotating the retro-reflection mirror of the atom interferometer, with no rotation applied to the sensor frame. These measurements are compared with the theoretical expression of contrast and phase shift given in Sec.~\ref{sec:level_1C}. Note that, compared to the previous section where the whole sensor is rotated, here we have the possibility to better control the rotation excitation and to implement a larger set of excitation configurations.

\begin{figure}
\centering
\includegraphics[scale=1]{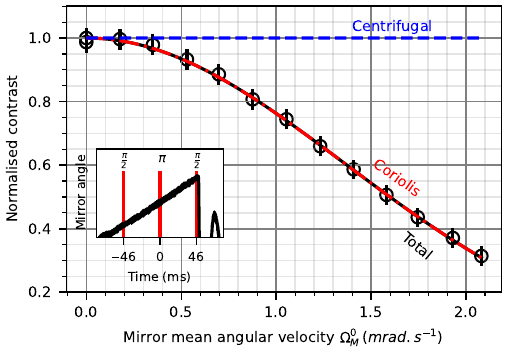} 
\caption{Impact of the mirror rotation on the interferometer contrast. The black line is the contrast as expected by Eq.~(\ref{eq_contrast_mirror}). In the inset, the black line is the mirror linear angular movement $\theta_M(t)$ measured by the capacitive detection. Legend similar to Fig.~\ref{fig_6_Contrast_Sensor_lin}.}
\label{fig_8_Contrast_mirror}
\end{figure}

\begin{figure}
\centering
\includegraphics[scale=1]{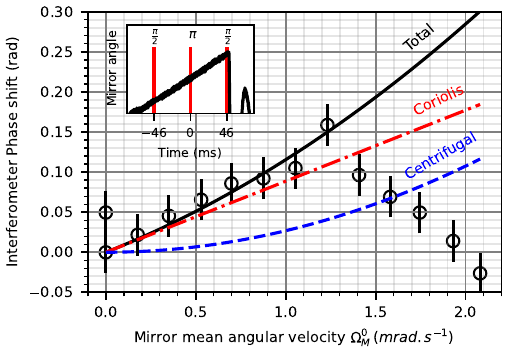} 
\caption{Impact of the mirror rotation on the interferometer phase shift. The black line is the phase shift as predicted by Eq.~(\ref{eq_phase_mirror}). The data are corrected from the residual vertical acceleration of the mirror. Legend similar to Fig.~\ref{fig_8_Contrast_mirror}.}
\label{fig_9_Phase_mirror}
\end{figure}

First, we study the simple case of an angular ramp applied to the mirror. Only Coriolis and centrifugal accelerations should contribute, as the mirror does not have any angular accelerations. As shown in Fig.~\ref{fig_8_Contrast_mirror}, the experimental contrast loss can only be explained by the contribution of the Coriolis acceleration. Similarly to Fig.~\ref{fig_6_Contrast_Sensor_lin}, the centrifugal acceleration is too small to be observed. On the whole data set, the theoretical model is in agreement with the experimental data of Fig.~\ref{fig_8_Contrast_mirror}. Note that the contrast loss induced by Coriolis acceleration is expected to be exactly the same in the case of a mirror rotation or in the case of the whole sensor rotation. In this section, the error bars associated with the mirror mean angular velocity are not visible. Nevertheless, the uncertainty was evaluated below \SI{3}{\micro \radian} resulting from the error on the EA angular capacitive detection reading and calibration.\\
In this simple case, the phase shift should be impacted only by the Coriolis and centrifugal accelerations. As can be seen in Fig.~\ref{fig_9_Phase_mirror}, the linear impact of Coriolis acceleration dominates for small mean angular velocities $\Omega^0_M$. The quadratic impact of the centrifugal acceleration is multiplied by two as the magnitude of the effective wavevector is modified by the mirror rotation (see Eq.~(\ref{eq_phase_mirror})). Moreover, the relevant distance for the mirror rotation is the vertical distance between the atom cloud and the center of rotation of the mirror, leading to a centrifugal term scaling as $2{\Omega^0_M}^2 \cdot \left[x_{MA} +T v_{x}+ a_x T^2\right]$. Due to the absence of angular acceleration, the experimental phase shift is smaller in this case. While the theoretical model is in agreement with the experimental data for angular velocities below $\approx$\SI{1}{\milli \radian \per \second}, some clear discrepancies up to \SI{0.3}{\radian} can be observed for higher angular velocities. This phase shift has not yet been fully understood and is still under investigation. According to a preliminary analysis, this behavior does not result from first-order spherical wavefront aberrations due to mirror surface imperfections as the EA proof-mass optical quality better than $\frac{\lambda}{4}$ (for $\lambda=\SI{600}{\nano \meter}$). Those anomalous measurements cannot be explained by a variation of the direction of measurement as the maximal mirror angle is of the order of \SI{100}{\micro \radian} leading to a phase shift of \SI{3}{\milli \radian}. Finally, this effect is still present when considering measurements that combine alternating signs of the laser wavevector  $\vec{k}_{\mathrm{eff}}$, ruling out one-photon light shift as the source of the effect.

\begin{figure*}
\centering
\includegraphics[scale=1]{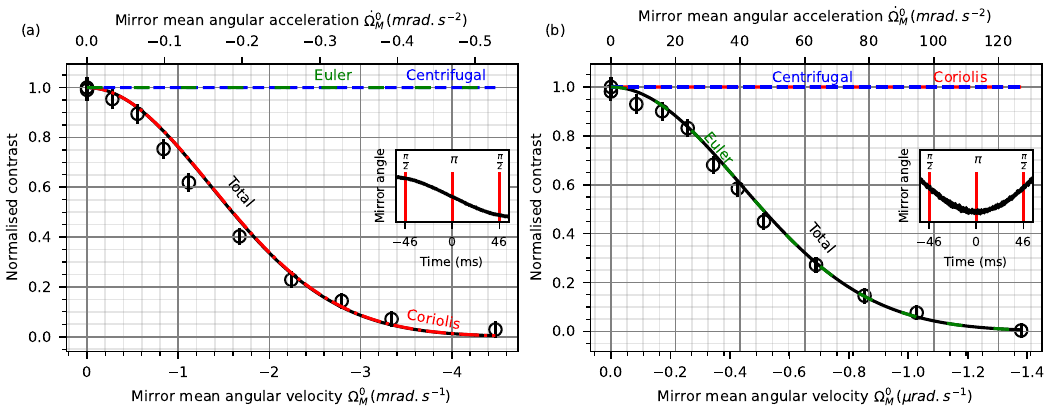}
\caption{Impact of the mirror rotation on the interferometer contrast. The black line is the theoretical model from Eq.~(\ref{eq_contrast_mirror}). In (a) (respectively (b)) the angular velocity is maximized (respectively minimized) and the angular acceleration is minimized (respectively maximized). Legend similar to Fig.~\ref{fig_8_Contrast_mirror}.}
\label{fig_10_contrast_mirror_sin}
\end{figure*}

Secondly, the effect of a sinusoidal angular movement of the mirror similar to the one presented in Section \ref{sec:level_4A} is studied. In this case, the Euler acceleration is minimized to study the Coriolis and centrifugal accelerations only. The range of accessible mean angular velocities is larger than for the sensor rotation, reaching a maximum angular velocity of \SI{4.48}{\milli \radian \per \second}. Moreover, the mean angular acceleration of the mirror remains below \SI{0.53}{\milli \radian \per \square \second} thanks to a better control of the angular movement. As can be seen in Fig.~\ref{fig_10_contrast_mirror_sin} (a) and as expected, the contrast loss is mainly due to the Coriolis acceleration, similarly to the mirror angular ramp case and to the sensor rotation, whereas the centrifugal and Euler accelerations play minor roles.

\begin{figure*}
\centering
\includegraphics[scale=1]{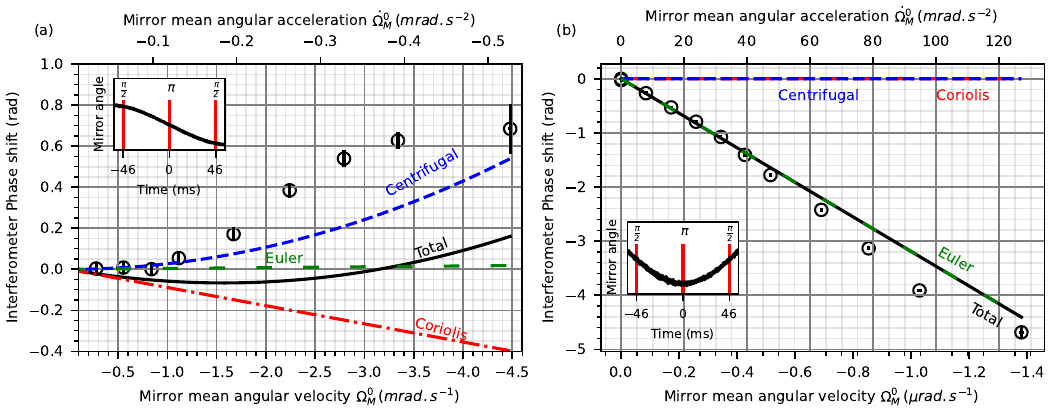}
\caption{Impact of the mirror rotation on the interferometer phase shift. The black line is the theoretical model from Eq.~(\ref{eq_phase_mirror}). The data are corrected from the residual vertical acceleration of the mirror. Legend similar to Fig.~\ref{fig_10_contrast_mirror_sin}.}
\label{fig_11_phase_mirror_sin}
\end{figure*}
The related phase shift (see Fig.~\ref{fig_11_phase_mirror_sin} (a)) is only affected by the Coriolis and centrifugal accelerations as the mirror rotation is better controlled than the sensor rotation, resulting in a smaller residual angular acceleration. Unfortunately, the phase shift shows large discrepancies up to \SI{0.6}{\radian} with the theoretical model which are not fully explained yet and are still under exploration, recalling the behavior previously described in the case of a linear angular ramp of the mirror.

We also studied configurations with important angular accelerations, maximizing the impact of the Euler acceleration on the interferometer output. In these configurations, the sinusoidal angular movement is phase shifted by $\pi/2$ compared to the previous case, inducing a maximum mean angular acceleration of \SI{127}{\milli \radian \per \square \second}. The mean angular velocity is minimized here and stays below \SI{1.4}{\micro \radian \per \second}. The only contribution to the contrast loss (see Fig.~\ref{fig_10_contrast_mirror_sin} (b)) is now due to the Euler acceleration,  linked to the velocity and position distribution of the atom cloud. Turning now to the impact on the phase shift (see Fig.~\ref{fig_11_phase_mirror_sin} (b)), we see clearly the high linear contribution, up to \SI{4.7}{\radian}, of the angular acceleration explained by the transverse distance between the atom cloud and the mirror center of rotation, resulting in the phase term $-k_{\mathrm{eff}} T^2\dot{\Omega}^0_M \cdot y_{MA}$. Deviations from the theoretical model seem in this case to be relatively much smaller when such high angular accelerations are generated. Still, for important rotation, some significant discrepancies up to \SI{0.63}{\radian} with the model appear but the experimental data stay in qualitative agreement. The errors to the theoretical model are of the same scale as the errors observed on Fig.~\ref{fig_9_Phase_mirror} and ~\ref{fig_11_phase_mirror_sin}(a) and might have the same origin.

\subsection{\label{sec:level_4C}Rotation compensation}

The rotation compensation method is implemented on the experimental setup by rotating the sensor according to a sinusoidal angular excitation at \SI{4}{\hertz} and, at the same time, rotating the mirror sinusoidally at the same frequency. The phase of the mirror excitation is adjusted experimentally so as to be in phase opposition compared to the sensor excitation. The ideal compensated rotation configuration corresponds to the situation where the mirror is not rotated anymore in the laboratory frame. For the following experimental results, the sensor rotation parameters are kept constant while the mirror rotation amplitude is scanned.

\begin{figure*}
\centering
\includegraphics[scale=1]{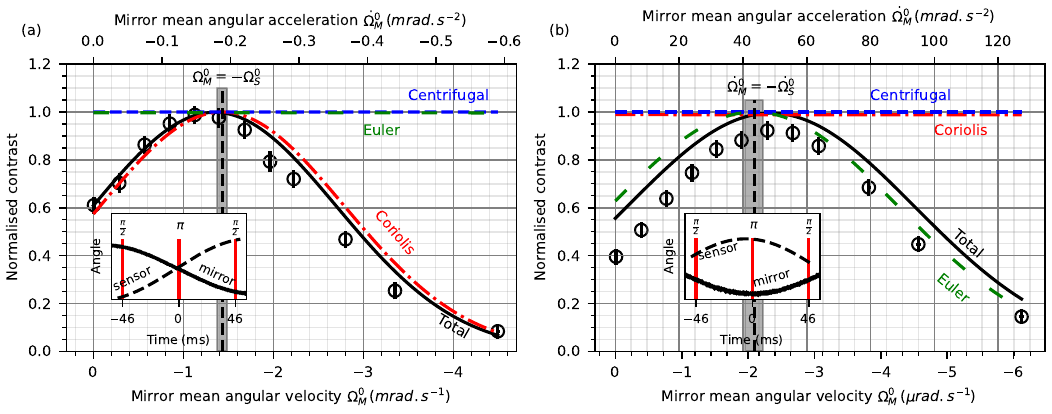}
\caption{Impact of the rotation compensation method on the interferometer contrast. The black line is the theoretical model from Eq.~(\ref{eq_contraste_compensation}). The gray area corresponds to the uncertainty on $\Omega_S^0$ in (a) (resp. $\dot{\Omega}_S^0$ in (b)). In the insets, the continuous black line is the mirror angular movement. The black dotted line is the sensor angular movement. In (a) (respectively (b)) the angular velocity is maximized (respectively minimized) and the angular acceleration is minimized (respectively maximized). Legend similar to Fig.~\ref{fig_10_contrast_mirror_sin}.}
\label{fig_12_contrast_compensation}
\end{figure*}

\begin{figure*}
\centering
\includegraphics[scale=1]{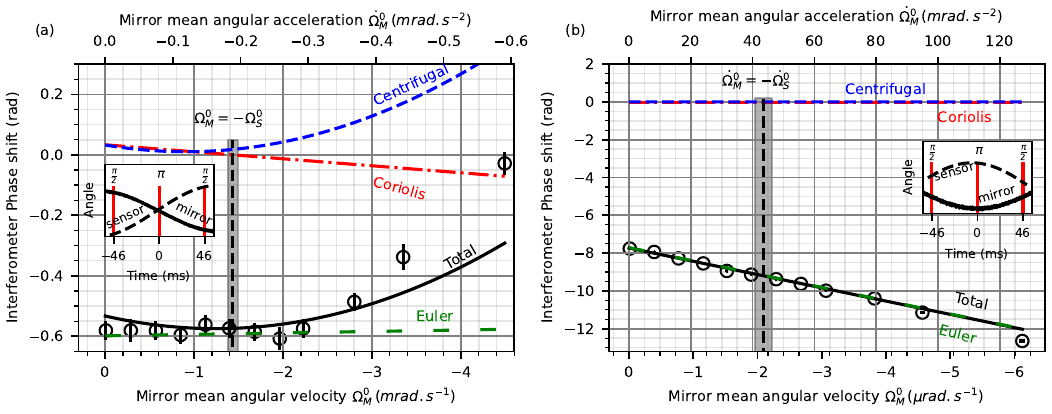}
\caption{Impact of the rotation compensation method on the interferometer phase shift. The black circles represent the experimental data with $1 \sigma$ error bars. The black line is the theoretical model from Eq.~(\ref{eq_phase_compensation_general}). Legend similar to Fig.~\ref{fig_12_contrast_compensation}.}
\label{fig_13_phase_compensation}
\end{figure*}

First, compensation of the angular velocity is studied with a minimized angular acceleration of the sensor, in order to focus on the correction of the effects induced by the Coriolis acceleration. For this study, the mean angular velocity of the sensor is set to $\Omega^0_S=\SI{1.43}{\milli \radian \per \second}$ with a minimized residual mean angular acceleration of the sensor $\dot{\Omega}^0_S=\SI{-3.5}{\milli \radian \per \square \second }$. The mirror is then rotated in the opposite direction in order to compensate for the Coriolis acceleration (see inset of Fig.~\ref{fig_12_contrast_compensation} (a)). The amplitude of the mirror rotation is scanned from an angular velocity of 0 to \SI{-4.49}{\milli \radian \per \second}. In Fig.~\ref{fig_12_contrast_compensation} (a), the contrast is gradually recovered as the mean angular velocity of the mirror increases, passing through a maximum close to $\Omega^0_M=-\Omega^0_S$, where the term $4(\Omega^0_M+\Omega^0_S)^2 \cdot \sigma_{v_y}^2$ cancels out. For higher mirror angular velocities, the sensor rotation is overcompensated, and the contrast decays. The compensation is successful as the contrast recovery reaches 99\%. As can be noticed, the contrast behavior is not fully driven by the Coriolis acceleration because of the residual Euler acceleration. This Euler acceleration is not compensated for by the mirror rotation and slightly shifts the optimal contrast recovery from $\Omega^0_M=-\Omega^0_S$.

With this compensation method, despite the fact that the impact of rotation on the phase shift cannot be fully canceled as predicted in Section \ref{sec:level_1C}, leaving a residual centrifugal acceleration term, a significant reduction of the phase shift could be expected. However, this is not experimentally observed, as can be seen in Fig.~\ref{fig_13_phase_compensation} (a) where the phase shift remains more or less constant around $\Omega^0_M \leq - \Omega_S$. Indeed, while the Coriolis acceleration is perfectly compensated and the centrifugal acceleration is reduced, an important uncompensated Euler acceleration dominates. As mentioned above, this is due to the fact that the residual mean Euler acceleration of the sensor rotation motion is not compensated for by the mirror rotation whose mean angular acceleration is much smaller. Nevertheless, around the compensation, the measured phase shift is retrieved by the model with a precision below \SI{0.025}{\radian}: the computed value could be useful to correct some future acceleration measurements. On the contrary, the theoretical model shows deviations up to \SI{0.3}{\radian} from the experimental data for mirror rotation higher than $\mid \Omega_M^0 \mid=\SI{2}{\milli \radian \per \s} $. An effect of the same magnitude was already observed in Section \ref{sec:level_4B}.

Secondly, we carry out experiments to study the compensation of the angular acceleration in order to cancel the effect of Euler acceleration. For that purpose, the sensor angular acceleration is maximized and set to $\dot{\Omega}^0_S=\SI{-43.5}{ \milli \radian \per \square \second}$, with a residual angular velocity $\Omega^0_S=\SI{-204}{\micro \radian \per \second}$. The mirror is rotated in the opposite way (see inset of Fig.~\ref{fig_12_contrast_compensation} (b)). The contrast is recovered as the rotation angular acceleration increases until it reaches a maximum of 92\% for $\dot{\Omega}^0_M=\SI{47.6}{\milli \radian \per \square \second}$ slightly shifted from the point $\dot{\Omega}^0_M=-\dot{\Omega}^0_S$. This effect can be explained by the presence of an uncompensated Coriolis acceleration in the atom velocity distribution term $ \left[ T (\dot{\Omega}^0_M+\dot{\Omega}^0_S) + 2 \Omega^0_S \right]^2 \sigma_{v_y}^2 $ assuming $\Omega^0_M=0$. Considering this residual Coriolis term, the optimal contrast recovery is expected for $\dot{\Omega}^0_M=\SI{52.4}{\milli \radian \per \square \second}$, close to the observed experimental value.

The phase shift induced by the Euler acceleration is not compensated as the centers of rotation of the mirror and the sensor do not coincide, leading to a residual term $-\dot{\Omega}^0_M \cdot y_{MO}$ (see Fig.~\ref{fig_13_phase_compensation} (b)). The centrifugal and Coriolis accelerations do not contribute to the phase shift as the mirror mean angular velocity is very small. In this configuration of high angular acceleration, the experimental results follow the global linear variation of the theoretical phase with the mirror mean angular acceleration. Nevertheless, an error to the model up to \SI{0.6}{\radian} is observed mostly for higher angular acceleration. At the compensation, the error is reduced to \SI{0.1}{\radian}. The differences between data and model could be explained by some variations in the distance $y_{MO}$ due to experimental imperfections in the rotation excitation of the sensor.
\section{\label{sec:level_5}Conclusion}
In this article, we present a detailed analysis of the impact of rotations on the phase and contrast of a cold-atom accelerometer. Analytical models are confronted to experimental results in different configurations of rotation excitation where the impact of angular velocity or angular acceleration is either minimized or maximized. The experimental study is being carried out using a unique hybrid atomic-electrostatic accelerometer where the atomic interferometer laser is retro-reflected on the proof-mass of an electrostatic accelerometer, similar to space geodesy missions instruments. This hybrid configuration allowed us to demonstrate experimentally the rotation compensation strategy of counter-rotating the interferometer mirror while rotating the whole instrument. We have shown that the contrast loss of the atomic interferometer is quite well reproduced by the theoretical model, regardless of the rotation excitation configuration in a range relevant for space geodesy applications. The use of an electrostatic accelerometer proof-mass as an actuated mirror enables the compensation of both angular velocities and accelerations. Contrast recoveries greater than 90\% were demonstrated in both cases. The results concerning the interferometer phase are not as well understood and are still under investigation. The phase shift data show deviations between the experimental data and the model, especially for mirror rotations of higher amplitudes.

This study is in line with previously published work that has aimed to analyze the detrimental impact of rotation on cold-atom accelerometers and has proposed mitigation strategies that would ultimately allow their operation in a dynamic environment \citep{beaufils_rotation_2023, darmagnac_de_castanet_atom_2024}. Our results, model, and experimental data are in agreement and bring additional and complementary insights by studying in detail the contribution of each non-inertial term, namely Coriolis, centrifugal, and Euler accelerations. We also propose an alternative to the common solution based on a PZT-actuated mirror. The benefits and drawbacks of each of these techniques have yet to be studied.

This topic is especially of primary importance in the perspective of future space missions on-boarding a cold-atom accelerometer \citep{Aguilera_2014, carraz_spaceborne_2014, chiow_laser-ranging_2015, abrykosov_impact_2019, migliaccio_mocass_2019, trimeche_concept_2019, leveque_gravity_2021, zahzam_hybrid_2022, hosseiniarani_advances_2024}, where satellite rotation could lead to detrimental issues. Our results constitute first experimental demonstrations that lead us to be confident in the possibility of retrieving the atomic contrast in orbit with the compensation method, considering that the atom source would be a delta-kick-collimated Bose-Einstein condensate \cite{hensel_inertial_2021} submitted to a Mach-Zehnder light pulse interferometer with an interrogation time $T \approx 1 s$. Concerning the impact of rotation on the interferometer phase, there is still a need in the near future to assess precisely the performance of the rotation compensation system, as it would directly impact the bias and stability of the measurement. Here we also initiate some discussion on the more appropriate position of a cold-atom instrument relative to the center-of-rotation of the satellite, so as to minimize rotation impact and propose a way to get rid of remaining centrifugal acceleration inherent to rotation compensation with mirror actuation. Finally, in this article, we also mention the potential of using the interferometer mirror rotation to characterize key parameters linked to the kinematics of the atom cloud that are of prime importance for the performance of an onboard atom interferometer. For instance, this type of characterization could be carried out during dedicated in-flight calibration phases.
%
%---------------------------------------------------------------------------------
\begin{acknowledgments} % Phuong-Anh Huynh (ONERA),
We thank Bernard Foulon (ONERA), Bruno Christophe (ONERA) and Fran\c{c}oise Liorzou (ONERA) for their early participation in the realization and characterization of the EA ground prototype. This work was partially supported by the European Space Agency (ESA) through the \textquotedblleft Hybrid Atom Electrostatic System Follow-On for Satellite Geodesy \textquotedblright, Contract No.4000122290/17/NL/FF/mg. The PhD grant of N. Marquet was co-funded by ESA and ONERA.
\end{acknowledgments}

%---------------------------------------------------------------------------------
%
\bibliography{Article_rotation}

%apsrev4-2.bst 2019-01-14 (MD) hand-edited version of apsrev4-1.bst
%Control: key (0)
%Control: author (72) initials jnrlst
%Control: editor formatted (1) identically to author
%Control: production of article title (-1) disabled
%Control: page (0) single
%Control: year (1) truncated
%Control: production of eprint (0) enabled
\begin{thebibliography}{48}%
\makeatletter
\providecommand \@ifxundefined [1]{%
 \@ifx{#1\undefined}
}%
\providecommand \@ifnum [1]{%
 \ifnum #1\expandafter \@firstoftwo
 \else \expandafter \@secondoftwo
 \fi
}%
\providecommand \@ifx [1]{%
 \ifx #1\expandafter \@firstoftwo
 \else \expandafter \@secondoftwo
 \fi
}%
\providecommand \natexlab [1]{#1}%
\providecommand \enquote  [1]{``#1''}%
\providecommand \bibnamefont  [1]{#1}%
\providecommand \bibfnamefont [1]{#1}%
\providecommand \citenamefont [1]{#1}%
\providecommand \href@noop [0]{\@secondoftwo}%
\providecommand \href [0]{\begingroup \@sanitize@url \@href}%
\providecommand \@href[1]{\@@startlink{#1}\@@href}%
\providecommand \@@href[1]{\endgroup#1\@@endlink}%
\providecommand \@sanitize@url [0]{\catcode `\\12\catcode `\$12\catcode
  `\&12\catcode `\#12\catcode `\^12\catcode `\_12\catcode `\%12\relax}%
\providecommand \@@startlink[1]{}%
\providecommand \@@endlink[0]{}%
\providecommand \url  [0]{\begingroup\@sanitize@url \@url }%
\providecommand \@url [1]{\endgroup\@href {#1}{\urlprefix }}%
\providecommand \urlprefix  [0]{URL }%
\providecommand \Eprint [0]{\href }%
\providecommand \doibase [0]{https://doi.org/}%
\providecommand \selectlanguage [0]{\@gobble}%
\providecommand \bibinfo  [0]{\@secondoftwo}%
\providecommand \bibfield  [0]{\@secondoftwo}%
\providecommand \translation [1]{[#1]}%
\providecommand \BibitemOpen [0]{}%
\providecommand \bibitemStop [0]{}%
\providecommand \bibitemNoStop [0]{.\EOS\space}%
\providecommand \EOS [0]{\spacefactor3000\relax}%
\providecommand \BibitemShut  [1]{\csname bibitem#1\endcsname}%
\let\auto@bib@innerbib\@empty
%</preamble>
\bibitem [{\citenamefont {Morel}\ \emph {et~al.}(2020)\citenamefont {Morel},
  \citenamefont {Yao}, \citenamefont {Cladé},\ and\ \citenamefont
  {Guellati-Khélifa}}]{morel_determination_2020}%
  \BibitemOpen
  \bibfield  {author} {\bibinfo {author} {\bibfnamefont {L.}~\bibnamefont
  {Morel}}, \bibinfo {author} {\bibfnamefont {Z.}~\bibnamefont {Yao}}, \bibinfo
  {author} {\bibfnamefont {P.}~\bibnamefont {Cladé}},\ and\ \bibinfo {author}
  {\bibfnamefont {S.}~\bibnamefont {Guellati-Khélifa}},\ }\href
  {https://doi.org/10.1038/s41586-020-2964-7} {\bibfield  {journal} {\bibinfo
  {journal} {Nature}\ }\textbf {\bibinfo {volume} {588}},\ \bibinfo {pages}
  {61} (\bibinfo {year} {2020})},\ \bibinfo {note} {publisher: Nature
  Publishing Group}\BibitemShut {NoStop}%
\bibitem [{\citenamefont {Parker}\ \emph {et~al.}(2018)\citenamefont {Parker},
  \citenamefont {Yu}, \citenamefont {Zhong}, \citenamefont {Estey},\ and\
  \citenamefont {Müller}}]{parker_measurement_2018}%
  \BibitemOpen
  \bibfield  {author} {\bibinfo {author} {\bibfnamefont {R.~H.}\ \bibnamefont
  {Parker}}, \bibinfo {author} {\bibfnamefont {C.}~\bibnamefont {Yu}}, \bibinfo
  {author} {\bibfnamefont {W.}~\bibnamefont {Zhong}}, \bibinfo {author}
  {\bibfnamefont {B.}~\bibnamefont {Estey}},\ and\ \bibinfo {author}
  {\bibfnamefont {H.}~\bibnamefont {Müller}},\ }\href
  {https://doi.org/10.1126/science.aap7706} {\bibfield  {journal} {\bibinfo
  {journal} {Science}\ }\textbf {\bibinfo {volume} {360}},\ \bibinfo {pages}
  {191} (\bibinfo {year} {2018})},\ \bibinfo {note} {publisher: American
  Association for the Advancement of Science}\BibitemShut {NoStop}%
\bibitem [{\citenamefont {Asenbaum}\ \emph {et~al.}(2020)\citenamefont
  {Asenbaum}, \citenamefont {Overstreet}, \citenamefont {Kim}, \citenamefont
  {Curti},\ and\ \citenamefont
  {Kasevich}}]{asenbaum_atom-interferometric_2020}%
  \BibitemOpen
  \bibfield  {author} {\bibinfo {author} {\bibfnamefont {P.}~\bibnamefont
  {Asenbaum}}, \bibinfo {author} {\bibfnamefont {C.}~\bibnamefont
  {Overstreet}}, \bibinfo {author} {\bibfnamefont {M.}~\bibnamefont {Kim}},
  \bibinfo {author} {\bibfnamefont {J.}~\bibnamefont {Curti}},\ and\ \bibinfo
  {author} {\bibfnamefont {M.~A.}\ \bibnamefont {Kasevich}},\ }\href
  {https://doi.org/10.1103/PhysRevLett.125.191101} {\bibfield  {journal}
  {\bibinfo  {journal} {Physical Review Letters}\ }\textbf {\bibinfo {volume}
  {125}},\ \bibinfo {pages} {191101} (\bibinfo {year} {2020})}\BibitemShut
  {NoStop}%
\bibitem [{\citenamefont {Jekeli}(2005)}]{jekeli_navigation_2005}%
  \BibitemOpen
  \bibfield  {author} {\bibinfo {author} {\bibfnamefont {C.}~\bibnamefont
  {Jekeli}},\ }\href
  {https://doi.org/https://doi.org/10.1002/j.2161-4296.2005.tb01726.x}
  {\bibfield  {journal} {\bibinfo  {journal} {Navigation}\ }\textbf {\bibinfo
  {volume} {52}},\ \bibinfo {pages} {1} (\bibinfo {year} {2005})}\BibitemShut
  {NoStop}%
\bibitem [{\citenamefont {Cheiney}\ \emph {et~al.}(2018)\citenamefont
  {Cheiney}, \citenamefont {Fouch\'e}, \citenamefont {Templier}, \citenamefont
  {Napolitano}, \citenamefont {Battelier}, \citenamefont {Bouyer},\ and\
  \citenamefont {Barrett}}]{cheiney_navigation_2018}%
  \BibitemOpen
  \bibfield  {author} {\bibinfo {author} {\bibfnamefont {P.}~\bibnamefont
  {Cheiney}}, \bibinfo {author} {\bibfnamefont {L.}~\bibnamefont {Fouch\'e}},
  \bibinfo {author} {\bibfnamefont {S.}~\bibnamefont {Templier}}, \bibinfo
  {author} {\bibfnamefont {F.}~\bibnamefont {Napolitano}}, \bibinfo {author}
  {\bibfnamefont {B.}~\bibnamefont {Battelier}}, \bibinfo {author}
  {\bibfnamefont {P.}~\bibnamefont {Bouyer}},\ and\ \bibinfo {author}
  {\bibfnamefont {B.}~\bibnamefont {Barrett}},\ }\href
  {https://doi.org/10.1103/PhysRevApplied.10.034030} {\bibfield  {journal}
  {\bibinfo  {journal} {Physical Review Applied}\ }\textbf {\bibinfo {volume}
  {10}},\ \bibinfo {pages} {034030} (\bibinfo {year} {2018})},\ \bibinfo {note}
  {publisher: American Physical Society}\BibitemShut {NoStop}%
\bibitem [{\citenamefont {Carraz}\ \emph {et~al.}(2014)\citenamefont {Carraz},
  \citenamefont {Siemes}, \citenamefont {Massotti}, \citenamefont {Haagmans},\
  and\ \citenamefont {Silvestrin}}]{carraz_spaceborne_2014}%
  \BibitemOpen
  \bibfield  {author} {\bibinfo {author} {\bibfnamefont {O.}~\bibnamefont
  {Carraz}}, \bibinfo {author} {\bibfnamefont {C.}~\bibnamefont {Siemes}},
  \bibinfo {author} {\bibfnamefont {L.}~\bibnamefont {Massotti}}, \bibinfo
  {author} {\bibfnamefont {R.}~\bibnamefont {Haagmans}},\ and\ \bibinfo
  {author} {\bibfnamefont {P.}~\bibnamefont {Silvestrin}},\ }\href
  {https://doi.org/10.1007/s12217-014-9385-x} {\bibfield  {journal} {\bibinfo
  {journal} {Microgravity Science and Technology}\ }\textbf {\bibinfo {volume}
  {26}},\ \bibinfo {pages} {139} (\bibinfo {year} {2014})}\BibitemShut
  {NoStop}%
\bibitem [{\citenamefont {Chiow}\ \emph {et~al.}(2015)\citenamefont {Chiow},
  \citenamefont {Williams},\ and\ \citenamefont
  {Yu}}]{chiow_laser-ranging_2015}%
  \BibitemOpen
  \bibfield  {author} {\bibinfo {author} {\bibfnamefont {S.-w.}\ \bibnamefont
  {Chiow}}, \bibinfo {author} {\bibfnamefont {J.}~\bibnamefont {Williams}},\
  and\ \bibinfo {author} {\bibfnamefont {N.}~\bibnamefont {Yu}},\ }\href
  {https://doi.org/10.1103/PhysRevA.92.063613} {\bibfield  {journal} {\bibinfo
  {journal} {Physical Review A}\ }\textbf {\bibinfo {volume} {92}},\ \bibinfo
  {pages} {063613} (\bibinfo {year} {2015})},\ \bibinfo {note} {publisher:
  American Physical Society}\BibitemShut {NoStop}%
\bibitem [{\citenamefont {Douch}\ \emph {et~al.}(2018)\citenamefont {Douch},
  \citenamefont {Wu}, \citenamefont {Schubert}, \citenamefont {M\"uller},\ and\
  \citenamefont {Pereira Dos~Santos}}]{douch_simulation-based_2018}%
  \BibitemOpen
  \bibfield  {author} {\bibinfo {author} {\bibfnamefont {K.}~\bibnamefont
  {Douch}}, \bibinfo {author} {\bibfnamefont {H.}~\bibnamefont {Wu}}, \bibinfo
  {author} {\bibfnamefont {C.}~\bibnamefont {Schubert}}, \bibinfo {author}
  {\bibfnamefont {J.}~\bibnamefont {M\"uller}},\ and\ \bibinfo {author}
  {\bibfnamefont {F.}~\bibnamefont {Pereira Dos~Santos}},\ }\href
  {https://doi.org/10.1016/j.asr.2017.12.005} {\bibfield  {journal} {\bibinfo
  {journal} {Advances in Space Research}\ }\textbf {\bibinfo {volume} {61}},\
  \bibinfo {pages} {1307} (\bibinfo {year} {2018})}\BibitemShut {NoStop}%
\bibitem [{\citenamefont {Trimeche}\ \emph {et~al.}(2019)\citenamefont
  {Trimeche}, \citenamefont {Battelier}, \citenamefont {Becker}, \citenamefont
  {Bertoldi}, \citenamefont {Bouyer}, \citenamefont {Braxmaier}, \citenamefont
  {Charron}, \citenamefont {Corgier}, \citenamefont {Cornelius}, \citenamefont
  {Douch}, \citenamefont {Gaaloul}, \citenamefont {Herrmann}, \citenamefont
  {Müller}, \citenamefont {Rasel}, \citenamefont {Schubert}, \citenamefont
  {Wu},\ and\ \citenamefont {Santos}}]{trimeche_concept_2019}%
  \BibitemOpen
  \bibfield  {author} {\bibinfo {author} {\bibfnamefont {A.}~\bibnamefont
  {Trimeche}}, \bibinfo {author} {\bibfnamefont {B.}~\bibnamefont {Battelier}},
  \bibinfo {author} {\bibfnamefont {D.}~\bibnamefont {Becker}}, \bibinfo
  {author} {\bibfnamefont {A.}~\bibnamefont {Bertoldi}}, \bibinfo {author}
  {\bibfnamefont {P.}~\bibnamefont {Bouyer}}, \bibinfo {author} {\bibfnamefont
  {C.}~\bibnamefont {Braxmaier}}, \bibinfo {author} {\bibfnamefont
  {E.}~\bibnamefont {Charron}}, \bibinfo {author} {\bibfnamefont
  {R.}~\bibnamefont {Corgier}}, \bibinfo {author} {\bibfnamefont
  {M.}~\bibnamefont {Cornelius}}, \bibinfo {author} {\bibfnamefont
  {K.}~\bibnamefont {Douch}}, \bibinfo {author} {\bibfnamefont
  {N.}~\bibnamefont {Gaaloul}}, \bibinfo {author} {\bibfnamefont
  {S.}~\bibnamefont {Herrmann}}, \bibinfo {author} {\bibfnamefont
  {J.}~\bibnamefont {Müller}}, \bibinfo {author} {\bibfnamefont
  {E.}~\bibnamefont {Rasel}}, \bibinfo {author} {\bibfnamefont
  {C.}~\bibnamefont {Schubert}}, \bibinfo {author} {\bibfnamefont
  {H.}~\bibnamefont {Wu}},\ and\ \bibinfo {author} {\bibfnamefont {F.~P.~d.}\
  \bibnamefont {Santos}},\ }\href {https://doi.org/10.1088/1361-6382/ab4548}
  {\bibfield  {journal} {\bibinfo  {journal} {Classical and Quantum Gravity}\
  }\textbf {\bibinfo {volume} {36}},\ \bibinfo {pages} {215004} (\bibinfo
  {year} {2019})},\ \bibinfo {note} {publisher: IOP Publishing}\BibitemShut
  {NoStop}%
\bibitem [{\citenamefont {Migliaccio}\ \emph {et~al.}(2019)\citenamefont
  {Migliaccio}, \citenamefont {Reguzzoni}, \citenamefont {Batsukh},
  \citenamefont {Tino}, \citenamefont {Rosi}, \citenamefont {Sorrentino},
  \citenamefont {Braitenberg}, \citenamefont {Pivetta}, \citenamefont
  {Barbolla},\ and\ \citenamefont {Zoffoli}}]{migliaccio_mocass_2019}%
  \BibitemOpen
  \bibfield  {author} {\bibinfo {author} {\bibfnamefont {F.}~\bibnamefont
  {Migliaccio}}, \bibinfo {author} {\bibfnamefont {M.}~\bibnamefont
  {Reguzzoni}}, \bibinfo {author} {\bibfnamefont {K.}~\bibnamefont {Batsukh}},
  \bibinfo {author} {\bibfnamefont {G.~M.}\ \bibnamefont {Tino}}, \bibinfo
  {author} {\bibfnamefont {G.}~\bibnamefont {Rosi}}, \bibinfo {author}
  {\bibfnamefont {F.}~\bibnamefont {Sorrentino}}, \bibinfo {author}
  {\bibfnamefont {C.}~\bibnamefont {Braitenberg}}, \bibinfo {author}
  {\bibfnamefont {T.}~\bibnamefont {Pivetta}}, \bibinfo {author} {\bibfnamefont
  {D.~F.}\ \bibnamefont {Barbolla}},\ and\ \bibinfo {author} {\bibfnamefont
  {S.}~\bibnamefont {Zoffoli}},\ }\href
  {https://doi.org/10.1007/s10712-019-09566-4} {\bibfield  {journal} {\bibinfo
  {journal} {Surveys in Geophysics}\ }\textbf {\bibinfo {volume} {40}},\
  \bibinfo {pages} {1029} (\bibinfo {year} {2019})}\BibitemShut {NoStop}%
\bibitem [{\citenamefont {Abrykosov}\ \emph {et~al.}(2019)\citenamefont
  {Abrykosov}, \citenamefont {Pail}, \citenamefont {Gruber}, \citenamefont
  {Zahzam}, \citenamefont {Bresson}, \citenamefont {Hardy}, \citenamefont
  {Christophe}, \citenamefont {Bidel}, \citenamefont {Carraz},\ and\
  \citenamefont {Siemes}}]{abrykosov_impact_2019}%
  \BibitemOpen
  \bibfield  {author} {\bibinfo {author} {\bibfnamefont {P.}~\bibnamefont
  {Abrykosov}}, \bibinfo {author} {\bibfnamefont {R.}~\bibnamefont {Pail}},
  \bibinfo {author} {\bibfnamefont {T.}~\bibnamefont {Gruber}}, \bibinfo
  {author} {\bibfnamefont {N.}~\bibnamefont {Zahzam}}, \bibinfo {author}
  {\bibfnamefont {A.}~\bibnamefont {Bresson}}, \bibinfo {author} {\bibfnamefont
  {E.}~\bibnamefont {Hardy}}, \bibinfo {author} {\bibfnamefont
  {B.}~\bibnamefont {Christophe}}, \bibinfo {author} {\bibfnamefont
  {Y.}~\bibnamefont {Bidel}}, \bibinfo {author} {\bibfnamefont
  {O.}~\bibnamefont {Carraz}},\ and\ \bibinfo {author} {\bibfnamefont
  {C.}~\bibnamefont {Siemes}},\ }\href
  {https://doi.org/10.1016/j.asr.2019.01.034} {\bibfield  {journal} {\bibinfo
  {journal} {Advances in Space Research}\ }\textbf {\bibinfo {volume} {63}},\
  \bibinfo {pages} {3235} (\bibinfo {year} {2019})}\BibitemShut {NoStop}%
\bibitem [{\citenamefont {L\'ev\`eque}\ \emph {et~al.}(2021)\citenamefont
  {L\'ev\`eque}, \citenamefont {Fallet}, \citenamefont {Mandea}, \citenamefont
  {Biancale}, \citenamefont {Lemoine}, \citenamefont {Tardivel}, \citenamefont
  {Delavault}, \citenamefont {Piquereau}, \citenamefont {Bourgogne},
  \citenamefont {Pereira Dos~Santos}, \citenamefont {Battelier},\ and\
  \citenamefont {Bouyer}}]{leveque_gravity_2021}%
  \BibitemOpen
  \bibfield  {author} {\bibinfo {author} {\bibfnamefont {T.}~\bibnamefont
  {L\'ev\`eque}}, \bibinfo {author} {\bibfnamefont {C.}~\bibnamefont {Fallet}},
  \bibinfo {author} {\bibfnamefont {M.}~\bibnamefont {Mandea}}, \bibinfo
  {author} {\bibfnamefont {R.}~\bibnamefont {Biancale}}, \bibinfo {author}
  {\bibfnamefont {J.~M.}\ \bibnamefont {Lemoine}}, \bibinfo {author}
  {\bibfnamefont {S.}~\bibnamefont {Tardivel}}, \bibinfo {author}
  {\bibfnamefont {S.}~\bibnamefont {Delavault}}, \bibinfo {author}
  {\bibfnamefont {A.}~\bibnamefont {Piquereau}}, \bibinfo {author}
  {\bibfnamefont {S.}~\bibnamefont {Bourgogne}}, \bibinfo {author}
  {\bibfnamefont {F.}~\bibnamefont {Pereira Dos~Santos}}, \bibinfo {author}
  {\bibfnamefont {B.}~\bibnamefont {Battelier}},\ and\ \bibinfo {author}
  {\bibfnamefont {P.}~\bibnamefont {Bouyer}},\ }\href
  {https://doi.org/10.1007/s00190-020-01462-9} {\bibfield  {journal} {\bibinfo
  {journal} {Journal of Geodesy}\ }\textbf {\bibinfo {volume} {95}},\ \bibinfo
  {pages} {15} (\bibinfo {year} {2021})}\BibitemShut {NoStop}%
\bibitem [{\citenamefont {Zahzam}\ \emph {et~al.}(2022)\citenamefont {Zahzam},
  \citenamefont {Christophe}, \citenamefont {Lebat}, \citenamefont {Hardy},
  \citenamefont {Huynh}, \citenamefont {Marquet}, \citenamefont {Blanchard},
  \citenamefont {Bidel}, \citenamefont {Bresson}, \citenamefont {Abrykosov},
  \citenamefont {Gruber}, \citenamefont {Pail}, \citenamefont {Daras},\ and\
  \citenamefont {Carraz}}]{zahzam_hybrid_2022}%
  \BibitemOpen
  \bibfield  {author} {\bibinfo {author} {\bibfnamefont {N.}~\bibnamefont
  {Zahzam}}, \bibinfo {author} {\bibfnamefont {B.}~\bibnamefont {Christophe}},
  \bibinfo {author} {\bibfnamefont {V.}~\bibnamefont {Lebat}}, \bibinfo
  {author} {\bibfnamefont {E.}~\bibnamefont {Hardy}}, \bibinfo {author}
  {\bibfnamefont {P.-A.}\ \bibnamefont {Huynh}}, \bibinfo {author}
  {\bibfnamefont {N.}~\bibnamefont {Marquet}}, \bibinfo {author} {\bibfnamefont
  {C.}~\bibnamefont {Blanchard}}, \bibinfo {author} {\bibfnamefont
  {Y.}~\bibnamefont {Bidel}}, \bibinfo {author} {\bibfnamefont
  {A.}~\bibnamefont {Bresson}}, \bibinfo {author} {\bibfnamefont
  {P.}~\bibnamefont {Abrykosov}}, \bibinfo {author} {\bibfnamefont
  {T.}~\bibnamefont {Gruber}}, \bibinfo {author} {\bibfnamefont
  {R.}~\bibnamefont {Pail}}, \bibinfo {author} {\bibfnamefont {I.}~\bibnamefont
  {Daras}},\ and\ \bibinfo {author} {\bibfnamefont {O.}~\bibnamefont
  {Carraz}},\ }\href {https://doi.org/10.3390/rs14143273} {\bibfield  {journal}
  {\bibinfo  {journal} {Remote Sensing}\ }\textbf {\bibinfo {volume} {14}},\
  \bibinfo {pages} {3273} (\bibinfo {year} {2022})},\ \bibinfo {note} {number:
  14 Publisher: Multidisciplinary Digital Publishing Institute}\BibitemShut
  {NoStop}%
\bibitem [{\citenamefont {HosseiniArani}\ \emph {et~al.}(2024)\citenamefont
  {HosseiniArani}, \citenamefont {Schilling}, \citenamefont {Beaufils},
  \citenamefont {Knabe}, \citenamefont {Tennstedt}, \citenamefont {Kupriyanov},
  \citenamefont {Schön}, \citenamefont {Pereira~dos Santos},\ and\
  \citenamefont {Müller}}]{hosseiniarani_advances_2024}%
  \BibitemOpen
  \bibfield  {author} {\bibinfo {author} {\bibfnamefont {A.}~\bibnamefont
  {HosseiniArani}}, \bibinfo {author} {\bibfnamefont {M.}~\bibnamefont
  {Schilling}}, \bibinfo {author} {\bibfnamefont {Q.}~\bibnamefont {Beaufils}},
  \bibinfo {author} {\bibfnamefont {A.}~\bibnamefont {Knabe}}, \bibinfo
  {author} {\bibfnamefont {B.}~\bibnamefont {Tennstedt}}, \bibinfo {author}
  {\bibfnamefont {A.}~\bibnamefont {Kupriyanov}}, \bibinfo {author}
  {\bibfnamefont {S.}~\bibnamefont {Schön}}, \bibinfo {author} {\bibfnamefont
  {F.}~\bibnamefont {Pereira~dos Santos}},\ and\ \bibinfo {author}
  {\bibfnamefont {J.}~\bibnamefont {Müller}},\ }\href
  {https://doi.org/10.1016/j.asr.2024.06.055} {\bibfield  {journal} {\bibinfo
  {journal} {Advances in Space Research}\ }\textbf {\bibinfo {volume} {74}},\
  \bibinfo {pages} {3186} (\bibinfo {year} {2024})}\BibitemShut {NoStop}%
\bibitem [{\citenamefont {Geiger}\ \emph {et~al.}(2011)\citenamefont {Geiger},
  \citenamefont {M\'enoret}, \citenamefont {Stern}, \citenamefont {Zahzam},
  \citenamefont {Cheinet}, \citenamefont {Battelier}, \citenamefont {Villing},
  \citenamefont {Moron}, \citenamefont {Lours}, \citenamefont {Bidel},
  \citenamefont {Bresson}, \citenamefont {Landragin},\ and\ \citenamefont
  {Bouyer}}]{geiger_detecting_2011}%
  \BibitemOpen
  \bibfield  {author} {\bibinfo {author} {\bibfnamefont {R.}~\bibnamefont
  {Geiger}}, \bibinfo {author} {\bibfnamefont {V.}~\bibnamefont {M\'enoret}},
  \bibinfo {author} {\bibfnamefont {G.}~\bibnamefont {Stern}}, \bibinfo
  {author} {\bibfnamefont {N.}~\bibnamefont {Zahzam}}, \bibinfo {author}
  {\bibfnamefont {P.}~\bibnamefont {Cheinet}}, \bibinfo {author} {\bibfnamefont
  {B.}~\bibnamefont {Battelier}}, \bibinfo {author} {\bibfnamefont
  {A.}~\bibnamefont {Villing}}, \bibinfo {author} {\bibfnamefont
  {F.}~\bibnamefont {Moron}}, \bibinfo {author} {\bibfnamefont
  {M.}~\bibnamefont {Lours}}, \bibinfo {author} {\bibfnamefont
  {Y.}~\bibnamefont {Bidel}}, \bibinfo {author} {\bibfnamefont
  {A.}~\bibnamefont {Bresson}}, \bibinfo {author} {\bibfnamefont
  {A.}~\bibnamefont {Landragin}},\ and\ \bibinfo {author} {\bibfnamefont
  {P.}~\bibnamefont {Bouyer}},\ }\href {https://doi.org/10.1038/ncomms1479}
  {\bibfield  {journal} {\bibinfo  {journal} {Nature Communications}\ }\textbf
  {\bibinfo {volume} {2}},\ \bibinfo {pages} {474} (\bibinfo {year} {2011})},\
  \bibinfo {note} {number: 1 Publisher: Nature Publishing Group}\BibitemShut
  {NoStop}%
\bibitem [{\citenamefont {Wu}\ \emph {et~al.}(2019)\citenamefont {Wu},
  \citenamefont {Pagel}, \citenamefont {Malek}, \citenamefont {Nguyen},
  \citenamefont {Zi}, \citenamefont {Scheirer},\ and\ \citenamefont
  {Müller}}]{wu_gravity_2019}%
  \BibitemOpen
  \bibfield  {author} {\bibinfo {author} {\bibfnamefont {X.}~\bibnamefont
  {Wu}}, \bibinfo {author} {\bibfnamefont {Z.}~\bibnamefont {Pagel}}, \bibinfo
  {author} {\bibfnamefont {B.~S.}\ \bibnamefont {Malek}}, \bibinfo {author}
  {\bibfnamefont {T.~H.}\ \bibnamefont {Nguyen}}, \bibinfo {author}
  {\bibfnamefont {F.}~\bibnamefont {Zi}}, \bibinfo {author} {\bibfnamefont
  {D.~S.}\ \bibnamefont {Scheirer}},\ and\ \bibinfo {author} {\bibfnamefont
  {H.}~\bibnamefont {Müller}},\ }\href
  {https://doi.org/10.1126/sciadv.aax0800} {\bibfield  {journal} {\bibinfo
  {journal} {Science Advances}\ }\textbf {\bibinfo {volume} {5}},\ \bibinfo
  {pages} {eaax0800} (\bibinfo {year} {2019})},\ \bibinfo {note} {publisher:
  American Association for the Advancement of Science}\BibitemShut {NoStop}%
\bibitem [{\citenamefont {Bidel}\ \emph {et~al.}(2018)\citenamefont {Bidel},
  \citenamefont {Zahzam}, \citenamefont {Blanchard}, \citenamefont {Bonnin},
  \citenamefont {Cadoret}, \citenamefont {Bresson}, \citenamefont {Rouxel},\
  and\ \citenamefont {Lequentrec-Lalancette}}]{bidel_absolute_2018}%
  \BibitemOpen
  \bibfield  {author} {\bibinfo {author} {\bibfnamefont {Y.}~\bibnamefont
  {Bidel}}, \bibinfo {author} {\bibfnamefont {N.}~\bibnamefont {Zahzam}},
  \bibinfo {author} {\bibfnamefont {C.}~\bibnamefont {Blanchard}}, \bibinfo
  {author} {\bibfnamefont {A.}~\bibnamefont {Bonnin}}, \bibinfo {author}
  {\bibfnamefont {M.}~\bibnamefont {Cadoret}}, \bibinfo {author} {\bibfnamefont
  {A.}~\bibnamefont {Bresson}}, \bibinfo {author} {\bibfnamefont
  {D.}~\bibnamefont {Rouxel}},\ and\ \bibinfo {author} {\bibfnamefont {M.~F.}\
  \bibnamefont {Lequentrec-Lalancette}},\ }\href
  {https://doi.org/10.1038/s41467-018-03040-2} {\bibfield  {journal} {\bibinfo
  {journal} {Nature Communications}\ }\textbf {\bibinfo {volume} {9}},\
  \bibinfo {pages} {627} (\bibinfo {year} {2018})},\ \bibinfo {note}
  {publisher: Nature Publishing Group}\BibitemShut {NoStop}%
\bibitem [{\citenamefont {Wu}\ \emph {et~al.}(2023)\citenamefont {Wu},
  \citenamefont {Zhang}, \citenamefont {Wang}, \citenamefont {Cheng},
  \citenamefont {Zhu}, \citenamefont {Li}, \citenamefont {Wang}, \citenamefont
  {Lin}, \citenamefont {Qiao},\ and\ \citenamefont {Zhou}}]{Wu_Marine_2023}%
  \BibitemOpen
  \bibfield  {author} {\bibinfo {author} {\bibfnamefont {B.}~\bibnamefont
  {Wu}}, \bibinfo {author} {\bibfnamefont {C.}~\bibnamefont {Zhang}}, \bibinfo
  {author} {\bibfnamefont {K.}~\bibnamefont {Wang}}, \bibinfo {author}
  {\bibfnamefont {B.}~\bibnamefont {Cheng}}, \bibinfo {author} {\bibfnamefont
  {D.}~\bibnamefont {Zhu}}, \bibinfo {author} {\bibfnamefont {R.}~\bibnamefont
  {Li}}, \bibinfo {author} {\bibfnamefont {X.}~\bibnamefont {Wang}}, \bibinfo
  {author} {\bibfnamefont {Q.}~\bibnamefont {Lin}}, \bibinfo {author}
  {\bibfnamefont {Z.}~\bibnamefont {Qiao}},\ and\ \bibinfo {author}
  {\bibfnamefont {Y.}~\bibnamefont {Zhou}},\ }\href
  {https://doi.org/10.1109/JSEN.2023.3309499} {\bibfield  {journal} {\bibinfo
  {journal} {IEEE Sensors Journal}\ }\textbf {\bibinfo {volume} {23}},\
  \bibinfo {pages} {24292} (\bibinfo {year} {2023})}\BibitemShut {NoStop}%
\bibitem [{\citenamefont {Bidel}\ \emph {et~al.}(2023)\citenamefont {Bidel},
  \citenamefont {Zahzam}, \citenamefont {Bresson}, \citenamefont {Blanchard},
  \citenamefont {Bonnin}, \citenamefont {Bernard}, \citenamefont {Cadoret},
  \citenamefont {Jensen}, \citenamefont {Forsberg}, \citenamefont {Salaun},
  \citenamefont {Lucas}, \citenamefont {Lequentrec-Lalancette}, \citenamefont
  {Rouxel}, \citenamefont {Gabalda}, \citenamefont {Seoane}, \citenamefont
  {Vu}, \citenamefont {Bruinsma},\ and\ \citenamefont
  {Bonvalot}}]{bidel_airborne_2023}%
  \BibitemOpen
  \bibfield  {author} {\bibinfo {author} {\bibfnamefont {Y.}~\bibnamefont
  {Bidel}}, \bibinfo {author} {\bibfnamefont {N.}~\bibnamefont {Zahzam}},
  \bibinfo {author} {\bibfnamefont {A.}~\bibnamefont {Bresson}}, \bibinfo
  {author} {\bibfnamefont {C.}~\bibnamefont {Blanchard}}, \bibinfo {author}
  {\bibfnamefont {A.}~\bibnamefont {Bonnin}}, \bibinfo {author} {\bibfnamefont
  {J.}~\bibnamefont {Bernard}}, \bibinfo {author} {\bibfnamefont
  {M.}~\bibnamefont {Cadoret}}, \bibinfo {author} {\bibfnamefont {T.~E.}\
  \bibnamefont {Jensen}}, \bibinfo {author} {\bibfnamefont {R.}~\bibnamefont
  {Forsberg}}, \bibinfo {author} {\bibfnamefont {C.}~\bibnamefont {Salaun}},
  \bibinfo {author} {\bibfnamefont {S.}~\bibnamefont {Lucas}}, \bibinfo
  {author} {\bibfnamefont {M.~F.}\ \bibnamefont {Lequentrec-Lalancette}},
  \bibinfo {author} {\bibfnamefont {D.}~\bibnamefont {Rouxel}}, \bibinfo
  {author} {\bibfnamefont {G.}~\bibnamefont {Gabalda}}, \bibinfo {author}
  {\bibfnamefont {L.}~\bibnamefont {Seoane}}, \bibinfo {author} {\bibfnamefont
  {D.~T.}\ \bibnamefont {Vu}}, \bibinfo {author} {\bibfnamefont
  {S.}~\bibnamefont {Bruinsma}},\ and\ \bibinfo {author} {\bibfnamefont
  {S.}~\bibnamefont {Bonvalot}},\ }\href {https://doi.org/10.1029/2022JB025921}
  {\bibfield  {journal} {\bibinfo  {journal} {Journal of Geophysical Research:
  Solid Earth}\ }\textbf {\bibinfo {volume} {128}},\ \bibinfo {pages}
  {e2022JB025921} (\bibinfo {year} {2023})}\BibitemShut {NoStop}%
\bibitem [{\citenamefont {Bongs}\ \emph {et~al.}(2006)\citenamefont {Bongs},
  \citenamefont {Launay},\ and\ \citenamefont
  {Kasevich}}]{bongs_high-order_2006}%
  \BibitemOpen
  \bibfield  {author} {\bibinfo {author} {\bibfnamefont {K.}~\bibnamefont
  {Bongs}}, \bibinfo {author} {\bibfnamefont {R.}~\bibnamefont {Launay}},\ and\
  \bibinfo {author} {\bibfnamefont {M.}~\bibnamefont {Kasevich}},\ }\href
  {https://doi.org/10.1007/s00340-006-2397-5} {\bibfield  {journal} {\bibinfo
  {journal} {Applied Physics B}\ }\textbf {\bibinfo {volume} {84}},\ \bibinfo
  {pages} {599} (\bibinfo {year} {2006})}\BibitemShut {NoStop}%
\bibitem [{\citenamefont {Hogan}\ \emph {et~al.}(2007)\citenamefont {Hogan},
  \citenamefont {Johnson},\ and\ \citenamefont
  {Kasevich}}]{hogan_light-pulse_2007}%
  \BibitemOpen
  \bibfield  {author} {\bibinfo {author} {\bibfnamefont {J.}~\bibnamefont
  {Hogan}}, \bibinfo {author} {\bibfnamefont {D.~M.~S.}\ \bibnamefont
  {Johnson}},\ and\ \bibinfo {author} {\bibfnamefont {M.}~\bibnamefont
  {Kasevich}},\ }\href
  {https://www.semanticscholar.org/paper/Light-pulse-atom-interferometry-Hogan-Johnson/fd05aca2c72f0a4d0c39584282ece346ea8450bd}
  {\bibfield  {journal} {\bibinfo  {journal} {Proceedings of the International
  Summer School of Physics "Enrico Fermi" on Atom Optics and Space Physics
  (Varenna)}\ } (\bibinfo {year} {2007})}\BibitemShut {NoStop}%
\bibitem [{\citenamefont {Lan}\ \emph {et~al.}(2012)\citenamefont {Lan},
  \citenamefont {Kuan}, \citenamefont {Estey}, \citenamefont {Haslinger},\ and\
  \citenamefont {Müller}}]{lan_influence_2012}%
  \BibitemOpen
  \bibfield  {author} {\bibinfo {author} {\bibfnamefont {S.-Y.}\ \bibnamefont
  {Lan}}, \bibinfo {author} {\bibfnamefont {P.-C.}\ \bibnamefont {Kuan}},
  \bibinfo {author} {\bibfnamefont {B.}~\bibnamefont {Estey}}, \bibinfo
  {author} {\bibfnamefont {P.}~\bibnamefont {Haslinger}},\ and\ \bibinfo
  {author} {\bibfnamefont {H.}~\bibnamefont {Müller}},\ }\href
  {https://doi.org/10.1103/PhysRevLett.108.090402} {\bibfield  {journal}
  {\bibinfo  {journal} {Physical Review Letters}\ }\textbf {\bibinfo {volume}
  {108}},\ \bibinfo {pages} {090402} (\bibinfo {year} {2012})}\BibitemShut
  {NoStop}%
\bibitem [{\citenamefont {Barrett}\ \emph {et~al.}(2016)\citenamefont
  {Barrett}, \citenamefont {Antoni-Micollier}, \citenamefont {Chichet},
  \citenamefont {Battelier}, \citenamefont {L\'ev\`eque}, \citenamefont
  {Landragin},\ and\ \citenamefont {Bouyer}}]{barrett_dual_2016}%
  \BibitemOpen
  \bibfield  {author} {\bibinfo {author} {\bibfnamefont {B.}~\bibnamefont
  {Barrett}}, \bibinfo {author} {\bibfnamefont {L.}~\bibnamefont
  {Antoni-Micollier}}, \bibinfo {author} {\bibfnamefont {L.}~\bibnamefont
  {Chichet}}, \bibinfo {author} {\bibfnamefont {B.}~\bibnamefont {Battelier}},
  \bibinfo {author} {\bibfnamefont {T.}~\bibnamefont {L\'ev\`eque}}, \bibinfo
  {author} {\bibfnamefont {A.}~\bibnamefont {Landragin}},\ and\ \bibinfo
  {author} {\bibfnamefont {P.}~\bibnamefont {Bouyer}},\ }\href
  {https://doi.org/10.1038/ncomms13786} {\bibfield  {journal} {\bibinfo
  {journal} {Nature Communications}\ }\textbf {\bibinfo {volume} {7}},\
  \bibinfo {pages} {13786} (\bibinfo {year} {2016})}\BibitemShut {NoStop}%
\bibitem [{\citenamefont {Zhao}\ \emph {et~al.}(2021)\citenamefont {Zhao},
  \citenamefont {Yue}, \citenamefont {Chen},\ and\ \citenamefont
  {Huang}}]{zhao_extension_2021}%
  \BibitemOpen
  \bibfield  {author} {\bibinfo {author} {\bibfnamefont {Y.}~\bibnamefont
  {Zhao}}, \bibinfo {author} {\bibfnamefont {X.}~\bibnamefont {Yue}}, \bibinfo
  {author} {\bibfnamefont {F.}~\bibnamefont {Chen}},\ and\ \bibinfo {author}
  {\bibfnamefont {C.}~\bibnamefont {Huang}},\ }\href
  {https://doi.org/10.1103/PhysRevA.104.013312} {\bibfield  {journal} {\bibinfo
   {journal} {Physical Review A}\ }\textbf {\bibinfo {volume} {104}},\ \bibinfo
  {pages} {013312} (\bibinfo {year} {2021})},\ \bibinfo {note} {publisher:
  American Physical Society}\BibitemShut {NoStop}%
\bibitem [{\citenamefont {Beaufils}\ \emph {et~al.}(2023)\citenamefont
  {Beaufils}, \citenamefont {Lefebve}, \citenamefont {Baptista}, \citenamefont
  {Piccon}, \citenamefont {Cambier}, \citenamefont {Sidorenkov}, \citenamefont
  {Fallet}, \citenamefont {Lévèque}, \citenamefont {Merlet},\ and\
  \citenamefont {Pereira Dos~Santos}}]{beaufils_rotation_2023}%
  \BibitemOpen
  \bibfield  {author} {\bibinfo {author} {\bibfnamefont {Q.}~\bibnamefont
  {Beaufils}}, \bibinfo {author} {\bibfnamefont {J.}~\bibnamefont {Lefebve}},
  \bibinfo {author} {\bibfnamefont {J.~G.}\ \bibnamefont {Baptista}}, \bibinfo
  {author} {\bibfnamefont {R.}~\bibnamefont {Piccon}}, \bibinfo {author}
  {\bibfnamefont {V.}~\bibnamefont {Cambier}}, \bibinfo {author} {\bibfnamefont
  {L.~A.}\ \bibnamefont {Sidorenkov}}, \bibinfo {author} {\bibfnamefont
  {C.}~\bibnamefont {Fallet}}, \bibinfo {author} {\bibfnamefont
  {T.}~\bibnamefont {Lévèque}}, \bibinfo {author} {\bibfnamefont
  {S.}~\bibnamefont {Merlet}},\ and\ \bibinfo {author} {\bibfnamefont
  {F.}~\bibnamefont {Pereira Dos~Santos}},\ }\href
  {https://doi.org/10.1038/s41526-023-00297-w} {\bibfield  {journal} {\bibinfo
  {journal} {npj Microgravity}\ }\textbf {\bibinfo {volume} {9}},\ \bibinfo
  {pages} {1} (\bibinfo {year} {2023})},\ \bibinfo {note} {publisher: Nature
  Publishing Group}\BibitemShut {NoStop}%
\bibitem [{\citenamefont {d’Armagnac~de Castanet}\ \emph
  {et~al.}(2024)\citenamefont {d’Armagnac~de Castanet}, \citenamefont
  {Des~Cognets}, \citenamefont {Arguel}, \citenamefont {Templier},
  \citenamefont {Jarlaud}, \citenamefont {Ménoret}, \citenamefont {Desruelle},
  \citenamefont {Bouyer},\ and\ \citenamefont
  {Battelier}}]{darmagnac_de_castanet_atom_2024}%
  \BibitemOpen
  \bibfield  {author} {\bibinfo {author} {\bibfnamefont {Q.}~\bibnamefont
  {d’Armagnac~de Castanet}}, \bibinfo {author} {\bibfnamefont
  {C.}~\bibnamefont {Des~Cognets}}, \bibinfo {author} {\bibfnamefont
  {R.}~\bibnamefont {Arguel}}, \bibinfo {author} {\bibfnamefont
  {S.}~\bibnamefont {Templier}}, \bibinfo {author} {\bibfnamefont
  {V.}~\bibnamefont {Jarlaud}}, \bibinfo {author} {\bibfnamefont
  {V.}~\bibnamefont {Ménoret}}, \bibinfo {author} {\bibfnamefont
  {B.}~\bibnamefont {Desruelle}}, \bibinfo {author} {\bibfnamefont
  {P.}~\bibnamefont {Bouyer}},\ and\ \bibinfo {author} {\bibfnamefont
  {B.}~\bibnamefont {Battelier}},\ }\href
  {https://doi.org/10.1038/s41467-024-50804-0} {\bibfield  {journal} {\bibinfo
  {journal} {Nature Communications}\ }\textbf {\bibinfo {volume} {15}},\
  \bibinfo {pages} {6406} (\bibinfo {year} {2024})},\ \bibinfo {note}
  {publisher: Nature Publishing Group}\BibitemShut {NoStop}%
\bibitem [{\citenamefont {Hauth}\ \emph {et~al.}(2013)\citenamefont {Hauth},
  \citenamefont {Freier}, \citenamefont {Schkolnik}, \citenamefont {Senger},
  \citenamefont {Schmidt},\ and\ \citenamefont {Peters}}]{hauth_first_2013}%
  \BibitemOpen
  \bibfield  {author} {\bibinfo {author} {\bibfnamefont {M.}~\bibnamefont
  {Hauth}}, \bibinfo {author} {\bibfnamefont {C.}~\bibnamefont {Freier}},
  \bibinfo {author} {\bibfnamefont {V.}~\bibnamefont {Schkolnik}}, \bibinfo
  {author} {\bibfnamefont {A.}~\bibnamefont {Senger}}, \bibinfo {author}
  {\bibfnamefont {M.}~\bibnamefont {Schmidt}},\ and\ \bibinfo {author}
  {\bibfnamefont {A.}~\bibnamefont {Peters}},\ }\href
  {https://doi.org/10.1007/s00340-013-5413-6} {\bibfield  {journal} {\bibinfo
  {journal} {Applied Physics B}\ }\textbf {\bibinfo {volume} {113}},\ \bibinfo
  {pages} {49} (\bibinfo {year} {2013})}\BibitemShut {NoStop}%
\bibitem [{\citenamefont {Dickerson}\ \emph {et~al.}(2013)\citenamefont
  {Dickerson}, \citenamefont {Hogan}, \citenamefont {Sugarbaker}, \citenamefont
  {Johnson},\ and\ \citenamefont {Kasevich}}]{dickerson_multiaxis_2013}%
  \BibitemOpen
  \bibfield  {author} {\bibinfo {author} {\bibfnamefont {S.~M.}\ \bibnamefont
  {Dickerson}}, \bibinfo {author} {\bibfnamefont {J.~M.}\ \bibnamefont
  {Hogan}}, \bibinfo {author} {\bibfnamefont {A.}~\bibnamefont {Sugarbaker}},
  \bibinfo {author} {\bibfnamefont {D.~M.~S.}\ \bibnamefont {Johnson}},\ and\
  \bibinfo {author} {\bibfnamefont {M.~A.}\ \bibnamefont {Kasevich}},\ }\href
  {https://doi.org/10.1103/PhysRevLett.111.083001} {\bibfield  {journal}
  {\bibinfo  {journal} {Physical Review Letters}\ }\textbf {\bibinfo {volume}
  {111}},\ \bibinfo {pages} {083001} (\bibinfo {year} {2013})},\ \bibinfo
  {note} {publisher: American Physical Society}\BibitemShut {NoStop}%
\bibitem [{\citenamefont {Duan}\ \emph {et~al.}(2020)\citenamefont {Duan},
  \citenamefont {He}, \citenamefont {Yan}, \citenamefont {Ji}, \citenamefont
  {Zhou}, \citenamefont {Chen}, \citenamefont {Wang},\ and\ \citenamefont
  {Zhan}}]{duan_suppression_2020}%
  \BibitemOpen
  \bibfield  {author} {\bibinfo {author} {\bibfnamefont {W.-T.}\ \bibnamefont
  {Duan}}, \bibinfo {author} {\bibfnamefont {C.}~\bibnamefont {He}}, \bibinfo
  {author} {\bibfnamefont {S.-T.}\ \bibnamefont {Yan}}, \bibinfo {author}
  {\bibfnamefont {Y.-H.}\ \bibnamefont {Ji}}, \bibinfo {author} {\bibfnamefont
  {L.}~\bibnamefont {Zhou}}, \bibinfo {author} {\bibfnamefont {X.}~\bibnamefont
  {Chen}}, \bibinfo {author} {\bibfnamefont {J.}~\bibnamefont {Wang}},\ and\
  \bibinfo {author} {\bibfnamefont {M.-S.}\ \bibnamefont {Zhan}},\ }\href
  {https://doi.org/10.1088/1674-1056/ab969a} {\bibfield  {journal} {\bibinfo
  {journal} {Chinese Physics B}\ }\textbf {\bibinfo {volume} {29}},\ \bibinfo
  {pages} {070305} (\bibinfo {year} {2020})}\BibitemShut {NoStop}%
\bibitem [{\citenamefont {Christophe}\ \emph {et~al.}(2015)\citenamefont
  {Christophe}, \citenamefont {Boulanger}, \citenamefont {Foulon},
  \citenamefont {Huynh}, \citenamefont {Lebat}, \citenamefont {Liorzou},\ and\
  \citenamefont {Perrot}}]{christophe_new_2015}%
  \BibitemOpen
  \bibfield  {author} {\bibinfo {author} {\bibfnamefont {B.}~\bibnamefont
  {Christophe}}, \bibinfo {author} {\bibfnamefont {D.}~\bibnamefont
  {Boulanger}}, \bibinfo {author} {\bibfnamefont {B.}~\bibnamefont {Foulon}},
  \bibinfo {author} {\bibfnamefont {P.~A.}\ \bibnamefont {Huynh}}, \bibinfo
  {author} {\bibfnamefont {V.}~\bibnamefont {Lebat}}, \bibinfo {author}
  {\bibfnamefont {F.}~\bibnamefont {Liorzou}},\ and\ \bibinfo {author}
  {\bibfnamefont {E.}~\bibnamefont {Perrot}},\ }\href
  {https://doi.org/10.1016/j.actaastro.2015.06.021} {\bibfield  {journal}
  {\bibinfo  {journal} {Acta Astronautica}\ }\textbf {\bibinfo {volume}
  {117}},\ \bibinfo {pages} {1} (\bibinfo {year} {2015})}\BibitemShut {NoStop}%
\bibitem [{\citenamefont {Kasevich}\ and\ \citenamefont
  {Chu}(1992)}]{kasevich_measurement_1992}%
  \BibitemOpen
  \bibfield  {author} {\bibinfo {author} {\bibfnamefont {M.}~\bibnamefont
  {Kasevich}}\ and\ \bibinfo {author} {\bibfnamefont {S.}~\bibnamefont {Chu}},\
  }\href {https://doi.org/10.1007/BF00325375} {\bibfield  {journal} {\bibinfo
  {journal} {Applied Physics B}\ }\textbf {\bibinfo {volume} {54}},\ \bibinfo
  {pages} {321} (\bibinfo {year} {1992})}\BibitemShut {NoStop}%
\bibitem [{\citenamefont {Bordé}(1989)}]{borde_atomic_1989}%
  \BibitemOpen
  \bibfield  {author} {\bibinfo {author} {\bibfnamefont {C.}~\bibnamefont
  {Bordé}},\ }\href@noop {} {\bibfield  {journal} {\bibinfo  {journal}
  {Physics Letters A}\ }\textbf {\bibinfo {volume} {140}},\ \bibinfo {pages}
  {10} (\bibinfo {year} {1989})}\BibitemShut {NoStop}%
\bibitem [{\citenamefont {Peters}\ \emph {et~al.}(2001)\citenamefont {Peters},
  \citenamefont {Chung},\ and\ \citenamefont
  {Chu}}]{peters_high-precision_2001}%
  \BibitemOpen
  \bibfield  {author} {\bibinfo {author} {\bibfnamefont {A.}~\bibnamefont
  {Peters}}, \bibinfo {author} {\bibfnamefont {K.~Y.}\ \bibnamefont {Chung}},\
  and\ \bibinfo {author} {\bibfnamefont {S.}~\bibnamefont {Chu}},\ }\href
  {https://doi.org/10.1088/0026-1394/38/1/4} {\bibfield  {journal} {\bibinfo
  {journal} {Metrologia}\ }\textbf {\bibinfo {volume} {38}},\ \bibinfo {pages}
  {25} (\bibinfo {year} {2001})}\BibitemShut {NoStop}%
\bibitem [{\citenamefont {Farah}\ \emph {et~al.}(2014)\citenamefont {Farah},
  \citenamefont {Gillot}, \citenamefont {Cheng}, \citenamefont {Landragin},
  \citenamefont {Merlet},\ and\ \citenamefont {Pereira
  Dos~Santos}}]{farah_effective_2014}%
  \BibitemOpen
  \bibfield  {author} {\bibinfo {author} {\bibfnamefont {T.}~\bibnamefont
  {Farah}}, \bibinfo {author} {\bibfnamefont {P.}~\bibnamefont {Gillot}},
  \bibinfo {author} {\bibfnamefont {B.}~\bibnamefont {Cheng}}, \bibinfo
  {author} {\bibfnamefont {A.}~\bibnamefont {Landragin}}, \bibinfo {author}
  {\bibfnamefont {S.}~\bibnamefont {Merlet}},\ and\ \bibinfo {author}
  {\bibfnamefont {F.}~\bibnamefont {Pereira Dos~Santos}},\ }\href
  {https://doi.org/10.1103/PhysRevA.90.023606} {\bibfield  {journal} {\bibinfo
  {journal} {Physical Review A}\ }\textbf {\bibinfo {volume} {90}},\ \bibinfo
  {pages} {023606} (\bibinfo {year} {2014})},\ \bibinfo {note} {publisher:
  American Physical Society}\BibitemShut {NoStop}%
\bibitem [{\citenamefont {Gillot}\ \emph {et~al.}(2016)\citenamefont {Gillot},
  \citenamefont {Cheng}, \citenamefont {Merlet},\ and\ \citenamefont {Pereira
  Dos~Santos}}]{gillot_limits_2016}%
  \BibitemOpen
  \bibfield  {author} {\bibinfo {author} {\bibfnamefont {P.}~\bibnamefont
  {Gillot}}, \bibinfo {author} {\bibfnamefont {B.}~\bibnamefont {Cheng}},
  \bibinfo {author} {\bibfnamefont {S.}~\bibnamefont {Merlet}},\ and\ \bibinfo
  {author} {\bibfnamefont {F.}~\bibnamefont {Pereira Dos~Santos}},\ }\href
  {https://doi.org/10.1103/PhysRevA.93.013609} {\bibfield  {journal} {\bibinfo
  {journal} {Physical Review A}\ }\textbf {\bibinfo {volume} {93}},\ \bibinfo
  {pages} {013609} (\bibinfo {year} {2016})},\ \bibinfo {note} {publisher:
  American Physical Society}\BibitemShut {NoStop}%
\bibitem [{\citenamefont {Werner}\ and\ \citenamefont
  {Wallis}(1993)}]{werner_laser_1993}%
  \BibitemOpen
  \bibfield  {author} {\bibinfo {author} {\bibfnamefont {J.}~\bibnamefont
  {Werner}}\ and\ \bibinfo {author} {\bibfnamefont {H.}~\bibnamefont
  {Wallis}},\ }\href {https://doi.org/10.1088/0953-4075/26/18/016} {\bibfield
  {journal} {\bibinfo  {journal} {Journal of Physics B: Atomic, Molecular and
  Optical Physics}\ }\textbf {\bibinfo {volume} {26}},\ \bibinfo {pages} {3063}
  (\bibinfo {year} {1993})}\BibitemShut {NoStop}%
\bibitem [{\citenamefont {Roura}(2017)}]{roura_circumventing_2017}%
  \BibitemOpen
  \bibfield  {author} {\bibinfo {author} {\bibfnamefont {A.}~\bibnamefont
  {Roura}},\ }\href {https://doi.org/10.1103/PhysRevLett.118.160401} {\bibfield
   {journal} {\bibinfo  {journal} {Phys. Rev. Lett.}\ }\textbf {\bibinfo
  {volume} {118}},\ \bibinfo {pages} {160401} (\bibinfo {year}
  {2017})}\BibitemShut {NoStop}%
\bibitem [{\citenamefont {D'Amico}\ \emph {et~al.}(2017)\citenamefont
  {D'Amico}, \citenamefont {Rosi}, \citenamefont {Zhan}, \citenamefont
  {Cacciapuoti}, \citenamefont {Fattori},\ and\ \citenamefont
  {Tino}}]{Damico_Cancel_2017}%
  \BibitemOpen
  \bibfield  {author} {\bibinfo {author} {\bibfnamefont {G.}~\bibnamefont
  {D'Amico}}, \bibinfo {author} {\bibfnamefont {G.}~\bibnamefont {Rosi}},
  \bibinfo {author} {\bibfnamefont {S.}~\bibnamefont {Zhan}}, \bibinfo {author}
  {\bibfnamefont {L.}~\bibnamefont {Cacciapuoti}}, \bibinfo {author}
  {\bibfnamefont {M.}~\bibnamefont {Fattori}},\ and\ \bibinfo {author}
  {\bibfnamefont {G.~M.}\ \bibnamefont {Tino}},\ }\href
  {https://doi.org/10.1103/PhysRevLett.119.253201} {\bibfield  {journal}
  {\bibinfo  {journal} {Phys. Rev. Lett.}\ }\textbf {\bibinfo {volume} {119}},\
  \bibinfo {pages} {253201} (\bibinfo {year} {2017})}\BibitemShut {NoStop}%
\bibitem [{\citenamefont {Overstreet}\ \emph {et~al.}(2018)\citenamefont
  {Overstreet}, \citenamefont {Asenbaum}, \citenamefont {Kovachy},
  \citenamefont {Notermans}, \citenamefont {Hogan},\ and\ \citenamefont
  {Kasevich}}]{Overstreet_Eff_2018}%
  \BibitemOpen
  \bibfield  {author} {\bibinfo {author} {\bibfnamefont {C.}~\bibnamefont
  {Overstreet}}, \bibinfo {author} {\bibfnamefont {P.}~\bibnamefont
  {Asenbaum}}, \bibinfo {author} {\bibfnamefont {T.}~\bibnamefont {Kovachy}},
  \bibinfo {author} {\bibfnamefont {R.}~\bibnamefont {Notermans}}, \bibinfo
  {author} {\bibfnamefont {J.~M.}\ \bibnamefont {Hogan}},\ and\ \bibinfo
  {author} {\bibfnamefont {M.~A.}\ \bibnamefont {Kasevich}},\ }\href
  {https://doi.org/10.1103/PhysRevLett.120.183604} {\bibfield  {journal}
  {\bibinfo  {journal} {Phys. Rev. Lett.}\ }\textbf {\bibinfo {volume} {120}},\
  \bibinfo {pages} {183604} (\bibinfo {year} {2018})}\BibitemShut {NoStop}%
\bibitem [{\citenamefont {Caldani}\ \emph {et~al.}(2019)\citenamefont
  {Caldani}, \citenamefont {Weng}, \citenamefont {Merlet},\ and\ \citenamefont
  {Pereira Dos~Santos}}]{Caldani_Simu_2019}%
  \BibitemOpen
  \bibfield  {author} {\bibinfo {author} {\bibfnamefont {R.}~\bibnamefont
  {Caldani}}, \bibinfo {author} {\bibfnamefont {K.~X.}\ \bibnamefont {Weng}},
  \bibinfo {author} {\bibfnamefont {S.}~\bibnamefont {Merlet}},\ and\ \bibinfo
  {author} {\bibfnamefont {F.}~\bibnamefont {Pereira Dos~Santos}},\ }\href
  {https://doi.org/10.1103/PhysRevA.99.033601} {\bibfield  {journal} {\bibinfo
  {journal} {Phys. Rev. A}\ }\textbf {\bibinfo {volume} {99}},\ \bibinfo
  {pages} {033601} (\bibinfo {year} {2019})}\BibitemShut {NoStop}%
\bibitem [{\citenamefont {Bidel}\ \emph {et~al.}(2013)\citenamefont {Bidel},
  \citenamefont {Carraz}, \citenamefont {Charrière}, \citenamefont {Cadoret},
  \citenamefont {Zahzam},\ and\ \citenamefont {Bresson}}]{bidel_compact_2013}%
  \BibitemOpen
  \bibfield  {author} {\bibinfo {author} {\bibfnamefont {Y.}~\bibnamefont
  {Bidel}}, \bibinfo {author} {\bibfnamefont {O.}~\bibnamefont {Carraz}},
  \bibinfo {author} {\bibfnamefont {R.}~\bibnamefont {Charrière}}, \bibinfo
  {author} {\bibfnamefont {M.}~\bibnamefont {Cadoret}}, \bibinfo {author}
  {\bibfnamefont {N.}~\bibnamefont {Zahzam}},\ and\ \bibinfo {author}
  {\bibfnamefont {A.}~\bibnamefont {Bresson}},\ }\href
  {https://doi.org/10.1063/1.4801756} {\bibfield  {journal} {\bibinfo
  {journal} {Applied Physics Letters}\ }\textbf {\bibinfo {volume} {102}},\
  \bibinfo {pages} {144107} (\bibinfo {year} {2013})}\BibitemShut {NoStop}%
\bibitem [{\citenamefont {Carraz}\ \emph {et~al.}(2009)\citenamefont {Carraz},
  \citenamefont {Lienhart}, \citenamefont {Charri{\`e}re}, \citenamefont
  {Cadoret}, \citenamefont {Zahzam},\ and\ \citenamefont
  {Bresson}}]{carraz_compact_2009}%
  \BibitemOpen
  \bibfield  {author} {\bibinfo {author} {\bibfnamefont {O.}~\bibnamefont
  {Carraz}}, \bibinfo {author} {\bibfnamefont {F.}~\bibnamefont {Lienhart}},
  \bibinfo {author} {\bibfnamefont {R.}~\bibnamefont {Charri{\`e}re}}, \bibinfo
  {author} {\bibfnamefont {M.}~\bibnamefont {Cadoret}}, \bibinfo {author}
  {\bibfnamefont {Y.}~\bibnamefont {Zahzam}, \bibfnamefont {N.and~Bidel}},\
  and\ \bibinfo {author} {\bibfnamefont {A.}~\bibnamefont {Bresson}},\
  }\href@noop {} {\bibfield  {journal} {\bibinfo  {journal} {Appl. Phys. B}\
  }\textbf {\bibinfo {volume} {97}},\ \bibinfo {pages} {405} (\bibinfo {year}
  {2009})}\BibitemShut {NoStop}%
\bibitem [{\citenamefont {Merlet}\ \emph {et~al.}(2009)\citenamefont {Merlet},
  \citenamefont {Gou\"et}, \citenamefont {Bodart}, \citenamefont {Clairon},
  \citenamefont {Landragin}, \citenamefont {Pereira Dos~Santos},\ and\
  \citenamefont {Rouchon}}]{merlet_operating_2009}%
  \BibitemOpen
  \bibfield  {author} {\bibinfo {author} {\bibfnamefont {S.}~\bibnamefont
  {Merlet}}, \bibinfo {author} {\bibfnamefont {J.~L.}\ \bibnamefont {Gou\"et}},
  \bibinfo {author} {\bibfnamefont {Q.}~\bibnamefont {Bodart}}, \bibinfo
  {author} {\bibfnamefont {A.}~\bibnamefont {Clairon}}, \bibinfo {author}
  {\bibfnamefont {A.}~\bibnamefont {Landragin}}, \bibinfo {author}
  {\bibfnamefont {F.}~\bibnamefont {Pereira Dos~Santos}},\ and\ \bibinfo
  {author} {\bibfnamefont {P.}~\bibnamefont {Rouchon}},\ }\href
  {https://doi.org/10.1088/0026-1394/46/1/011} {\bibfield  {journal} {\bibinfo
  {journal} {Metrologia}\ }\textbf {\bibinfo {volume} {46}},\ \bibinfo {pages}
  {87} (\bibinfo {year} {2009})},\ \bibinfo {note} {publisher: IOP
  Publishing}\BibitemShut {NoStop}%
\bibitem [{\citenamefont {Lautier}\ \emph {et~al.}(2014)\citenamefont
  {Lautier}, \citenamefont {Volodimer}, \citenamefont {Hardin}, \citenamefont
  {Merlet}, \citenamefont {Lours}, \citenamefont {Pereira Dos~Santos},\ and\
  \citenamefont {Landragin}}]{lautier_hybridizing_2014}%
  \BibitemOpen
  \bibfield  {author} {\bibinfo {author} {\bibfnamefont {J.}~\bibnamefont
  {Lautier}}, \bibinfo {author} {\bibfnamefont {L.}~\bibnamefont {Volodimer}},
  \bibinfo {author} {\bibfnamefont {T.}~\bibnamefont {Hardin}}, \bibinfo
  {author} {\bibfnamefont {S.}~\bibnamefont {Merlet}}, \bibinfo {author}
  {\bibfnamefont {M.}~\bibnamefont {Lours}}, \bibinfo {author} {\bibfnamefont
  {F.}~\bibnamefont {Pereira Dos~Santos}},\ and\ \bibinfo {author}
  {\bibfnamefont {A.}~\bibnamefont {Landragin}},\ }\href
  {https://doi.org/10.1063/1.4897358} {\bibfield  {journal} {\bibinfo
  {journal} {Applied Physics Letters}\ }\textbf {\bibinfo {volume} {105}},\
  \bibinfo {pages} {144102} (\bibinfo {year} {2014})},\ \bibinfo {note}
  {publisher: American Institute of Physics}\BibitemShut {NoStop}%
\bibitem [{\citenamefont {Rodrigues}\ \emph {et~al.}(2022)\citenamefont
  {Rodrigues}, \citenamefont {Bergé}, \citenamefont {Boulanger}, \citenamefont
  {Christophe}, \citenamefont {Dalin}, \citenamefont {Lebat},\ and\
  \citenamefont {Liorzou}}]{rodrigues_space_2022}%
  \BibitemOpen
  \bibfield  {author} {\bibinfo {author} {\bibfnamefont {M.}~\bibnamefont
  {Rodrigues}}, \bibinfo {author} {\bibfnamefont {J.}~\bibnamefont {Bergé}},
  \bibinfo {author} {\bibfnamefont {D.}~\bibnamefont {Boulanger}}, \bibinfo
  {author} {\bibfnamefont {B.}~\bibnamefont {Christophe}}, \bibinfo {author}
  {\bibfnamefont {M.}~\bibnamefont {Dalin}}, \bibinfo {author} {\bibfnamefont
  {V.}~\bibnamefont {Lebat}},\ and\ \bibinfo {author} {\bibfnamefont
  {F.}~\bibnamefont {Liorzou}},\ }\href@noop {} {\bibfield  {journal} {\bibinfo
   {journal} {The 4S Symposium 2022}\ } (\bibinfo {year} {2022})}\BibitemShut
  {NoStop}%
\bibitem [{\citenamefont {Bidel}\ \emph {et~al.}(2020)\citenamefont {Bidel},
  \citenamefont {Zahzam}, \citenamefont {Bresson}, \citenamefont {Blanchard},
  \citenamefont {Cadoret}, \citenamefont {Olesen},\ and\ \citenamefont
  {Forsberg}}]{bidel_absolute_2020}%
  \BibitemOpen
  \bibfield  {author} {\bibinfo {author} {\bibfnamefont {Y.}~\bibnamefont
  {Bidel}}, \bibinfo {author} {\bibfnamefont {N.}~\bibnamefont {Zahzam}},
  \bibinfo {author} {\bibfnamefont {A.}~\bibnamefont {Bresson}}, \bibinfo
  {author} {\bibfnamefont {C.}~\bibnamefont {Blanchard}}, \bibinfo {author}
  {\bibfnamefont {M.}~\bibnamefont {Cadoret}}, \bibinfo {author} {\bibfnamefont
  {A.~V.}\ \bibnamefont {Olesen}},\ and\ \bibinfo {author} {\bibfnamefont
  {R.}~\bibnamefont {Forsberg}},\ }\href
  {https://doi.org/10.1007/s00190-020-01350-2} {\bibfield  {journal} {\bibinfo
  {journal} {Journal of Geodesy}\ }\textbf {\bibinfo {volume} {94}},\ \bibinfo
  {pages} {20} (\bibinfo {year} {2020})}\BibitemShut {NoStop}%
\bibitem [{\citenamefont {Aguilera}\ \emph {et~al.}(2014)\citenamefont
  {Aguilera}, \citenamefont {Ahlers}, \citenamefont {Battelier}, \citenamefont
  {Bawamia}, \citenamefont {Bertoldi}, \citenamefont {Bondarescu},
  \citenamefont {Bongs}, \citenamefont {Bouyer}, \citenamefont {Braxmaier},
  \citenamefont {Cacciapuoti}, \citenamefont {Chaloner}, \citenamefont
  {Chwalla}, \citenamefont {Ertmer}, \citenamefont {Franz}, \citenamefont
  {Gaaloul}, \citenamefont {Gehler}, \citenamefont {Gerardi}, \citenamefont
  {Gesa}, \citenamefont {Gürlebeck}, \citenamefont {Hartwig}, \citenamefont
  {Hauth}, \citenamefont {Hellmig}, \citenamefont {Herr}, \citenamefont
  {Herrmann}, \citenamefont {Heske}, \citenamefont {Hinton}, \citenamefont
  {Ireland}, \citenamefont {Jetzer}, \citenamefont {Johann}, \citenamefont
  {Krutzik}, \citenamefont {Kubelka}, \citenamefont {Lämmerzahl},
  \citenamefont {Landragin}, \citenamefont {Lloro}, \citenamefont {Massonnet},
  \citenamefont {Mateos}, \citenamefont {Milke}, \citenamefont {Nofrarias},
  \citenamefont {Oswald}, \citenamefont {Peters}, \citenamefont
  {Posso-Trujillo}, \citenamefont {Rasel}, \citenamefont {Rocco}, \citenamefont
  {Roura}, \citenamefont {Rudolph}, \citenamefont {Schleich}, \citenamefont
  {Schubert}, \citenamefont {Schuldt}, \citenamefont {Seidel}, \citenamefont
  {Sengstock}, \citenamefont {Sopuerta}, \citenamefont {Sorrentino},
  \citenamefont {Summers}, \citenamefont {Tino}, \citenamefont {Trenkel},
  \citenamefont {Uzunoglu}, \citenamefont {von Klitzing}, \citenamefont
  {Walser}, \citenamefont {Wendrich}, \citenamefont {Wenzlawski}, \citenamefont
  {Weßels}, \citenamefont {Wicht}, \citenamefont {Wille}, \citenamefont
  {Williams}, \citenamefont {Windpassinger},\ and\ \citenamefont
  {Zahzam}}]{Aguilera_2014}%
  \BibitemOpen
  \bibfield  {author} {\bibinfo {author} {\bibfnamefont {D.~N.}\ \bibnamefont
  {Aguilera}}, \bibinfo {author} {\bibfnamefont {H.}~\bibnamefont {Ahlers}},
  \bibinfo {author} {\bibfnamefont {B.}~\bibnamefont {Battelier}}, \bibinfo
  {author} {\bibfnamefont {A.}~\bibnamefont {Bawamia}}, \bibinfo {author}
  {\bibfnamefont {A.}~\bibnamefont {Bertoldi}}, \bibinfo {author}
  {\bibfnamefont {R.}~\bibnamefont {Bondarescu}}, \bibinfo {author}
  {\bibfnamefont {K.}~\bibnamefont {Bongs}}, \bibinfo {author} {\bibfnamefont
  {P.}~\bibnamefont {Bouyer}}, \bibinfo {author} {\bibfnamefont
  {C.}~\bibnamefont {Braxmaier}}, \bibinfo {author} {\bibfnamefont
  {L.}~\bibnamefont {Cacciapuoti}}, \bibinfo {author} {\bibfnamefont
  {C.}~\bibnamefont {Chaloner}}, \bibinfo {author} {\bibfnamefont
  {M.}~\bibnamefont {Chwalla}}, \bibinfo {author} {\bibfnamefont
  {W.}~\bibnamefont {Ertmer}}, \bibinfo {author} {\bibfnamefont
  {M.}~\bibnamefont {Franz}}, \bibinfo {author} {\bibfnamefont
  {N.}~\bibnamefont {Gaaloul}}, \bibinfo {author} {\bibfnamefont
  {M.}~\bibnamefont {Gehler}}, \bibinfo {author} {\bibfnamefont
  {D.}~\bibnamefont {Gerardi}}, \bibinfo {author} {\bibfnamefont
  {L.}~\bibnamefont {Gesa}}, \bibinfo {author} {\bibfnamefont {N.}~\bibnamefont
  {Gürlebeck}}, \bibinfo {author} {\bibfnamefont {J.}~\bibnamefont {Hartwig}},
  \bibinfo {author} {\bibfnamefont {M.}~\bibnamefont {Hauth}}, \bibinfo
  {author} {\bibfnamefont {O.}~\bibnamefont {Hellmig}}, \bibinfo {author}
  {\bibfnamefont {W.}~\bibnamefont {Herr}}, \bibinfo {author} {\bibfnamefont
  {S.}~\bibnamefont {Herrmann}}, \bibinfo {author} {\bibfnamefont
  {A.}~\bibnamefont {Heske}}, \bibinfo {author} {\bibfnamefont
  {A.}~\bibnamefont {Hinton}}, \bibinfo {author} {\bibfnamefont
  {P.}~\bibnamefont {Ireland}}, \bibinfo {author} {\bibfnamefont
  {P.}~\bibnamefont {Jetzer}}, \bibinfo {author} {\bibfnamefont
  {U.}~\bibnamefont {Johann}}, \bibinfo {author} {\bibfnamefont
  {M.}~\bibnamefont {Krutzik}}, \bibinfo {author} {\bibfnamefont
  {A.}~\bibnamefont {Kubelka}}, \bibinfo {author} {\bibfnamefont
  {C.}~\bibnamefont {Lämmerzahl}}, \bibinfo {author} {\bibfnamefont
  {A.}~\bibnamefont {Landragin}}, \bibinfo {author} {\bibfnamefont
  {I.}~\bibnamefont {Lloro}}, \bibinfo {author} {\bibfnamefont
  {D.}~\bibnamefont {Massonnet}}, \bibinfo {author} {\bibfnamefont
  {I.}~\bibnamefont {Mateos}}, \bibinfo {author} {\bibfnamefont
  {A.}~\bibnamefont {Milke}}, \bibinfo {author} {\bibfnamefont
  {M.}~\bibnamefont {Nofrarias}}, \bibinfo {author} {\bibfnamefont
  {M.}~\bibnamefont {Oswald}}, \bibinfo {author} {\bibfnamefont
  {A.}~\bibnamefont {Peters}}, \bibinfo {author} {\bibfnamefont
  {K.}~\bibnamefont {Posso-Trujillo}}, \bibinfo {author} {\bibfnamefont
  {E.}~\bibnamefont {Rasel}}, \bibinfo {author} {\bibfnamefont
  {E.}~\bibnamefont {Rocco}}, \bibinfo {author} {\bibfnamefont
  {A.}~\bibnamefont {Roura}}, \bibinfo {author} {\bibfnamefont
  {J.}~\bibnamefont {Rudolph}}, \bibinfo {author} {\bibfnamefont
  {W.}~\bibnamefont {Schleich}}, \bibinfo {author} {\bibfnamefont
  {C.}~\bibnamefont {Schubert}}, \bibinfo {author} {\bibfnamefont
  {T.}~\bibnamefont {Schuldt}}, \bibinfo {author} {\bibfnamefont
  {S.}~\bibnamefont {Seidel}}, \bibinfo {author} {\bibfnamefont
  {K.}~\bibnamefont {Sengstock}}, \bibinfo {author} {\bibfnamefont {C.~F.}\
  \bibnamefont {Sopuerta}}, \bibinfo {author} {\bibfnamefont {F.}~\bibnamefont
  {Sorrentino}}, \bibinfo {author} {\bibfnamefont {D.}~\bibnamefont {Summers}},
  \bibinfo {author} {\bibfnamefont {G.~M.}\ \bibnamefont {Tino}}, \bibinfo
  {author} {\bibfnamefont {C.}~\bibnamefont {Trenkel}}, \bibinfo {author}
  {\bibfnamefont {N.}~\bibnamefont {Uzunoglu}}, \bibinfo {author}
  {\bibfnamefont {W.}~\bibnamefont {von Klitzing}}, \bibinfo {author}
  {\bibfnamefont {R.}~\bibnamefont {Walser}}, \bibinfo {author} {\bibfnamefont
  {T.}~\bibnamefont {Wendrich}}, \bibinfo {author} {\bibfnamefont
  {A.}~\bibnamefont {Wenzlawski}}, \bibinfo {author} {\bibfnamefont
  {P.}~\bibnamefont {Weßels}}, \bibinfo {author} {\bibfnamefont
  {A.}~\bibnamefont {Wicht}}, \bibinfo {author} {\bibfnamefont
  {E.}~\bibnamefont {Wille}}, \bibinfo {author} {\bibfnamefont
  {M.}~\bibnamefont {Williams}}, \bibinfo {author} {\bibfnamefont
  {P.}~\bibnamefont {Windpassinger}},\ and\ \bibinfo {author} {\bibfnamefont
  {N.}~\bibnamefont {Zahzam}},\ }\href
  {https://doi.org/10.1088/0264-9381/31/11/115010} {\bibfield  {journal}
  {\bibinfo  {journal} {Classical and Quantum Gravity}\ }\textbf {\bibinfo
  {volume} {31}},\ \bibinfo {pages} {115010} (\bibinfo {year}
  {2014})}\BibitemShut {NoStop}%
\bibitem [{\citenamefont {Hensel}\ \emph {et~al.}(2021)\citenamefont {Hensel},
  \citenamefont {Loriani}, \citenamefont {Schubert}, \citenamefont {Fitzek},
  \citenamefont {Abend}, \citenamefont {Ahlers}, \citenamefont {Siemß},
  \citenamefont {Hammerer}, \citenamefont {Rasel},\ and\ \citenamefont
  {Gaaloul}}]{hensel_inertial_2021}%
  \BibitemOpen
  \bibfield  {author} {\bibinfo {author} {\bibfnamefont {T.}~\bibnamefont
  {Hensel}}, \bibinfo {author} {\bibfnamefont {S.}~\bibnamefont {Loriani}},
  \bibinfo {author} {\bibfnamefont {C.}~\bibnamefont {Schubert}}, \bibinfo
  {author} {\bibfnamefont {F.}~\bibnamefont {Fitzek}}, \bibinfo {author}
  {\bibfnamefont {S.}~\bibnamefont {Abend}}, \bibinfo {author} {\bibfnamefont
  {H.}~\bibnamefont {Ahlers}}, \bibinfo {author} {\bibfnamefont {J.-N.}\
  \bibnamefont {Siemß}}, \bibinfo {author} {\bibfnamefont {K.}~\bibnamefont
  {Hammerer}}, \bibinfo {author} {\bibfnamefont {E.~M.}\ \bibnamefont
  {Rasel}},\ and\ \bibinfo {author} {\bibfnamefont {N.}~\bibnamefont
  {Gaaloul}},\ }\href {https://doi.org/10.1140/epjd/s10053-021-00069-9}
  {\bibfield  {journal} {\bibinfo  {journal} {The European Physical Journal D}\
  }\textbf {\bibinfo {volume} {75}},\ \bibinfo {pages} {108} (\bibinfo {year}
  {2021})}\BibitemShut {NoStop}%
\end{thebibliography}%

\newpage
\appendix
\section{Acceleration of the mirror}
Due to limits in the control of the actuated mirror, a vertical movement was observed during the rotation experiments. The variation of the EA position with respect to a rest position $\delta x_{EA}$ along the $\vec{x}_S$ axis is measured during the interferometer by use of capacitive detection. As the interferometer is sensitive to the mirror acceleration, the induced phase shift can be computed with Eq.~(\ref{eq_phase_pistonEA}).\\
\begin{equation}\label{eq_phase_pistonEA}
\begin{split}
\Delta \Phi_{acc_{EA}} &=k_{\mathrm{eff}}T^2 \left[ \right. \delta x_{EA}(t_0)-2 \delta x_{EA}(t_0+T)\\
&+\delta x_{EA}(t_0+2T) \left] \right.
\end{split}
\end{equation}
\section{Two-axis rotation of the sensor}
In contrast with what is expected, the sensor does not rotate purely around the $\vec{z}_S$ axis. A small component around the $\vec{y}_S$ axis was measured with the two-axis gyroscope. The induced effect on the phase shift was computed with Eq.~(\ref{eq_phase_sensor_y}) and subtracted from the experimental data presented in the article.\\
\begin{equation}\label{eq_phase_sensor_y}
\begin{split}
\overline{\Delta \Phi}_{Sy}&=k_{\mathrm{eff}} T^2 \Bigg[ \Bigg. 2\Omega^0_{Sy} \cdot \left (v_z + \frac{3}{2} T a_z \right) \\
&+\dot{\Omega}^0_{Sy} \cdot (z_{OA}+ v_z T + \frac{1}{2} T^2 a_z)\\
&+{\Omega^0_{Sy}}^2 \cdot \left(x_{OA} + 3 v_x T + \frac{7}{2} T^2 a_x \right) \Bigg] \Bigg.
\end{split}
\end{equation}
with $\Omega^0_{Sy}$ and $\dot{\Omega}^0_{Sy}$ the mean angular velocity and mean angular acceleration around $\vec{y}_S$ during the interferometer acquisition sequence.
\section{Rotating the mirror, a way to characterize the atom cloud}
The rotation of the mirror was used to measure the kinematic parameters of the atom cloud at launch (index 0 on parameters) such as its temperature, size, mean velocity, and position. To do so, the phase shift and contrast are now expressed as a function of the parameters at launch.\\

\begin{figure}
\centering
\includegraphics[scale=1]{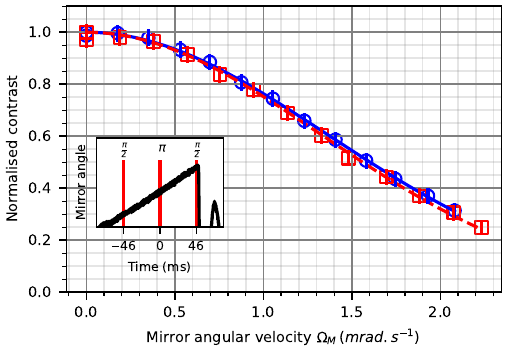} 
\caption{Temperature measurements through the Coriolis induced contrast loss. The blue circles (respectively the red squares) are the experimental data for a rotation around the $\vec{z}_M$ axis (respectively the $\vec{y}_M$ axis). The plain blue line (resp. the dotted red line) is the fit curve of the experimental data by Eq.~(\ref{eq_contraste_temperature}). The angular movement of the mirror, black line in the inset, is a ramp measured by the capacitive detection. The red lines represent the laser pulses of the interferometer.}
\label{fig_14_contrast_temperature}
\end{figure}

\noindent
\textbf{Velocity distribution:} The width of the velocity distribution was measured through the contrast loss induced by rotation. In the absence of angular acceleration, the contrast loss is only due to Coriolis and centrifugal acceleration. The contribution of the latter is negligible in this study.\\
\begin{equation}\label{eq_contraste_temperature}
\begin{split}
\frac{\overline{C}}{C}&=\exp\Bigg[ - 2 k_{\mathrm{eff}}^2 T^4 {\Omega^0_M}^2 \sigma_{v_i}^2 \Bigg]
\end{split}
\end{equation}
An angular ramp was imposed to the mirror around the $\vec{z}_M$ axis to measure the velocity distribution along the $\vec{y}_M$ axis and vice versa (see Fig.~\ref{fig_14_contrast_temperature}). The normalized contrast was fitted with Eq.~(\ref{eq_contraste_temperature}) with $i \in \lbrace y;z \rbrace$. The width of the velocity distribution was measured at $\sigma_{v_y}=\SI{10.8(2)}{\mm \per \s} $ along the $\vec{y}_M$ axis and $\sigma_{v_z}=\SI{11.1(2)}{\mm \per \s}$ along the $\vec{z}_M$ axis.\\

\noindent
\textbf{Size:} The same principle was applied to measure the size of the atom cloud right after the cooling stage. In the absence of an angular velocity, the contrast loss is due only to the angular acceleration and the velocity and position distributions of the cloud. As the velocity distribution was already measured, the position distribution can be determined.\\

\begin{figure}
\centering
\includegraphics[scale=1]{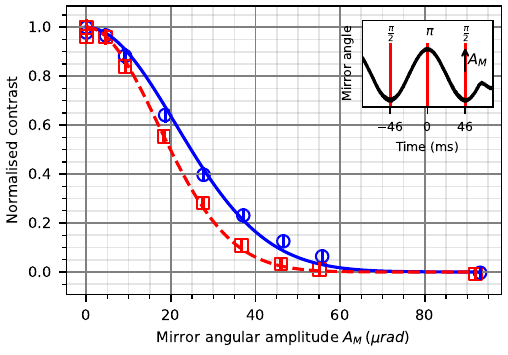} 
\caption{Size measurements through the Euler induced contrast loss. The plain blue line (resp. the dotted red line) is the fit curve of the experimental data by Eq.~(\ref{eq_contraste_position}). The angular movement of the mirror is a sine as described in the inset. Legend similar to Fig.~\ref{fig_14_contrast_temperature}.}
\label{fig_15_contrast_size}
\end{figure}
\begin{equation}\label{eq_contraste_position}
\begin{split}
\frac{\overline{C}}{C}&=\exp\Bigg[ - 8 k_{\mathrm{eff}}^2 A_M^2 \cdot \Bigg( \sigma_{i}^2 +(t_0+T)^2 \sigma_{v_i}^2 \Bigg) \Bigg]
\end{split}
\end{equation}
A sine movement of angular amplitude $A_M$ was imposed to the mirror around the $\vec{z}_M$ axis to measure the size along the $\vec{y}_M$ axis and vice versa (see Fig.~\ref{fig_15_contrast_size}). Such an angular movement maximizes the angular acceleration of the mirror. The normalized contrast was fitted with Eq.~(\ref{eq_contraste_position}) with $i \in \lbrace y;z \rbrace$. The width of the position distribution was measured at $\sigma_{y_0}=\SI{0.42(5)}{\-mm}$ along the $\vec{y}_M$ axis and $\sigma_{z_0}=\SI{0.66(5)}{\mm}$ along the $\vec{z}_M$ axis.\\

\begin{figure}
\centering
\includegraphics[scale=1]{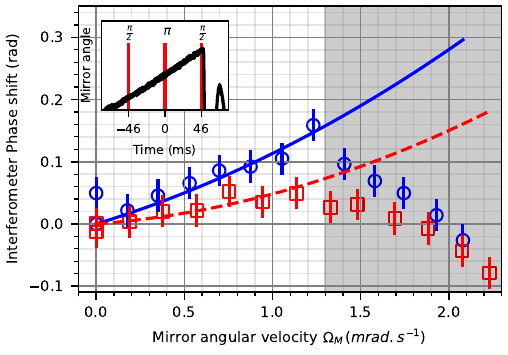} 
\caption{Mean velocity measurements through the Coriolis induced phase shift. The plain blue line (resp. the dotted red line) is the fit curve of the experimental data by Eq.~(\ref{eq_phase_vitesseZ}) (resp. Eq.~(\ref{eq_phase_vitesseY})). The gray area delimits the excluded points from the fit. The angular movement of the mirror is a ramp as described in the inset. Legend similar to Fig.~\ref{fig_14_contrast_temperature}.}
\label{fig_16_phase_velocity}
\end{figure}

\noindent
\textbf{Mean velocity:} The mean velocity of the atom cloud was measured through the Coriolis-induced phase shift. The impact of centrifugal acceleration on the phase shift cannot be ignored but can be estimated as the vertical kinematic parameters of the cloud are known.
\begin{equation}\label{eq_phase_vitesseZ}
\begin{split}
\Delta \Phi &= k_{\mathrm{eff}} T^2 \Bigg[ \Bigg. -2\Omega_M \cdot v_{y0} \\
&+ 2{\Omega_M}^2(a_{x_L} \cdot (\frac{t_0^2}{2}+t_0T+T^2)\\
&+v_{x_0}(t_0+T)+x_{MA}) \Bigg] \Bigg.\\
\end{split}
\end{equation}
\begin{equation}\label{eq_phase_vitesseY}
\begin{split}
\Delta \Phi &= k_{\mathrm{eff}} T^2 \Bigg[ \Bigg. 2\Omega_M \cdot v_{z0} \\
&+ 2{\Omega_M}^2(a_{x_L} \cdot (\frac{t_0^2}{2}+t_0T+T^2)\\
&+v_{x_0}(t_0+T)+x_{MA}) \Bigg] \Bigg.\\
\end{split}
\end{equation}
An angular ramp was imposed to the mirror around the $\vec{z}_M$ axis to measure the velocity along the $\vec{y}_M$ axis and vice versa (see Fig.~\ref{fig_16_phase_velocity}). The phase shift was fitted with Eq.~(\ref{eq_phase_vitesseZ}) for the rotation along the $\vec{z}_M$ axis (in blue) and with Eq.~(\ref{eq_phase_vitesseY}) for the rotation along the $\vec{y}_M$ axis (in red). The mean velocity was measured at $v_{y_0}=\SI{-1.3(3)}{\mm \per \s} $ along the $\vec{y}_M$ axis and $v_{z_0}=\SI{0.3(3)}{\mm \per \s} $ along the $\vec{z}_M$ axis.\\

\begin{figure}
\centering
\includegraphics[scale=1]{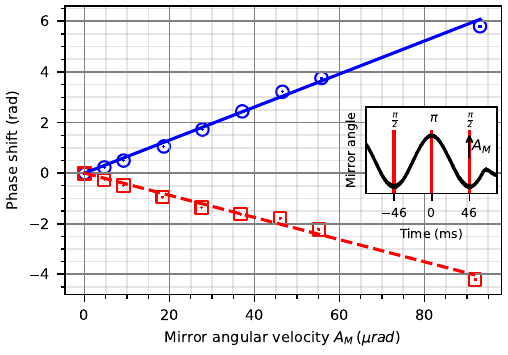} 
\caption{Mean position measurements through the Euler induced phase shift. The plain blue line (resp. the dotted red line) is the fit curve of the experimental data by Eq.~(\ref{eq_phase_positionZ}) (resp. Eq.~(\ref{eq_phase_positionY})). The angular movement of the mirror is a sine as described in the inset. Legend similar to Fig.~\ref{fig_14_contrast_temperature}}
\label{fig_17_phase_position}
\end{figure}

\noindent
\textbf{Mean position:} The same principle was applied to measure the mean position of the atom cloud right after the cooling stage. In the absence of an angular velocity, the phase shift is due only to the Euler acceleration and the velocity and position of the cloud. As the velocity was already measured, the position can be determined.\\

\begin{equation}\label{eq_phase_positionZ}
\Delta \Phi = k_{\mathrm{eff}} T^2 \left[ \frac{4A_M}{T^2} \cdot \left(  (t_0+T)v_{y_0}+y_{MA} \right) \right]
\end{equation}
\begin{equation}\label{eq_phase_positionY}
\Delta \Phi = k_{\mathrm{eff}} T^2 \left[ - \frac{4A_M}{T^2} \cdot \left(  (t_0+T)v_{z_0}+z_{MA} \right)  \right] 
\end{equation}
A sine movement was imposed to the mirror around the $\vec{z}_M$ axis to measure the position along the $\vec{y}_M$ axis and vice versa (see Fig.~\ref{fig_17_phase_position}). The phase shift was fitted with Eq.~(\ref{eq_phase_positionZ}) for the rotation around the $\vec{z}_M$ axis (in blue) and with Eq.~(\ref{eq_phase_positionY}) for the rotation around the $\vec{y}_M$ axis (in red). The mean position was measured at $y_{MA}=\SI{1.09(3)}{\mm} $ along the $\vec{y}_M$ axis and $z_{MA}=\SI{0.66(3)}{\mm} $ along the $\vec{z}_M$ axis.\\

\end{document}